\documentclass[%
 amsmath,amssymb,
 aps,
pra,
longbibliography,
notitlepage,
]{revtex4-1}
\UseRawInputEncoding
\usepackage{graphicx}% Include figure files
\usepackage{dcolumn}% Align table columns on decimal point
\usepackage{bm}% bold math
\usepackage{tikz}
\usetikzlibrary{shapes}
\usepackage{color}
\usepackage{float}
\usepackage{placeins}
\usepackage[caption=false]{subfig}
\definecolor{magenta}{rgb}{0.79216,0.12156,0.48236}

\newcommand{\bw}{\mbox{$\overline{w}$} {}}
\newcommand{\bv}{\mbox{$\bar{v}_r$} {}}

\newcommand{\uu}{\mbox{\boldmath $u$} {}}
\newcommand{\ww}{\mbox{\boldmath $w$} {}}
\newcommand{\ee}{\mbox{\boldmath $e$} {}}
\newcommand{\vv}{\mbox{\boldmath $v$} {}}
\newcommand{\xx}{\mbox{\boldmath $x$} {}}

\newcommand{\ttau}{\mbox{\boldmath $\tau$} {}}
\newcommand{\ppi}{\mbox{\boldmath $\pi$} {}}
\newcommand{\tuu}{\mbox{\boldmath $\tilde{u}$} {}}
\newcommand{\tvv}{\mbox{\boldmath $\tilde{v}$} {}}

\newcommand{\tp}{\mbox{$\tilde{P}$} {}}

\newcommand{\tn}{\mbox{$\tilde{n}$} {}}
\newcommand{\tS}{\mbox{$\tilde{S}$} {}}

\newcommand{\overbar}[1]{\mkern 1.5mu\overline{\mkern-1.5mu#1\mkern-1.5mu}\mkern 1.5mu}

\newcommand{\blackline}{\raisebox{2pt}{\tikz{\draw[-,black,solid,line width = 0.9pt](0,0) -- (5mm,0);}}}
\newcommand{\blackdline}{\raisebox{2pt}{\tikz{\draw[-,black,dashed,line width = 0.9pt](0,0) -- (5mm,0);}}}
\newcommand{\blackldline}{\raisebox{2pt}{\tikz{\draw[-,black,dash pattern=on 3pt off 6pt,line width = 0.9pt](0,0) -- (5mm,0);}}}
\newcommand{\reddline}{\raisebox{2pt}{\tikz{\draw[-,red,dashed,line width = 0.9pt](0,0) -- (5mm,0);}}}
\newcommand{\redline}{\raisebox{2pt}{\tikz{\draw[-,red,solid,line width = 1.5pt](0,0) -- (5mm,0);}}}
\newcommand{\blueline}{\raisebox{2pt}{\tikz{\draw[-,blue,solid,line width = 1.5pt](0,0) -- (5mm,0);}}}
\newcommand{\blueddline}{\raisebox{2pt}{\tikz{\draw[-,blue,dotted,line width = 1.5pt](0,0) -- (5mm,0);}}}

\newcommand{\greendline}{\raisebox{2pt}{\tikz{\draw[-,black!60!green,dashed,line width = 1.5pt](0,0) -- (5mm,0);}}}
\newcommand{\magline}{\raisebox{2pt}{\tikz{\draw[-,magenta,solid,line width = 1.5pt](0,0) -- (5mm,0);}}}
\newcommand{\magdline}{\raisebox{2pt}{\tikz{\draw[-,magenta,dash pattern={on 5pt off 1pt on 2pt off 1pt on 5pt},line width = 1.5pt](0,0) -- (5mm,0);}}}
\newcommand{\circleline}{\raisebox{0pt}{\tikz{\draw[black,solid,line width = 1.0pt,fill=none](2.8mm,0.8mm) circle (0.8mm);\draw[-,black,solid,line width = 1.0pt](0.,0.8mm) -- (5.5mm,0.8mm)}}}

\newcommand{\sqline}{\raisebox{0pt}{\tikz{\draw[black,solid,line width = 1.0pt,fill=none](2mm,0mm) rectangle (3.5mm,1.5mm);\draw[-,black,solid,line width = 1.0pt](0.,0.8mm) -- (5.5mm,0.8mm)}}}

\newcommand{\rtri}{\raisebox{0.5pt}{\tikz{\node[draw,red,scale=0.4,regular polygon, regular polygon sides=3,fill=none](){};}}}
\newcommand{\mtri}{\textcolor{magenta}{\rhd}}
\newcommand{\sq}{\raisebox{0.5pt}{\tikz{\node[draw,red,scale=0.6,regular polygon, regular polygon sides=4,fill=none](){};}}}
\newcommand{\bsq}{\raisebox{0.5pt}{\tikz{\node[draw,blue,scale=0.6,regular polygon, regular polygon sides=4,fill=none](){};}}}
\newcommand{\kcircle}{\raisebox{0.5pt}{\tikz{\node[draw,black,scale=0.4,circle,line width = 1.0pt,fill=none](){};}}}
\newcommand{\rcircle}{\raisebox{0.5pt}{\tikz{\node[draw,red,scale=0.4,circle,line width = 1.0pt,fill=none](){};}}}
\newcommand{\bcircle}{\raisebox{0.5pt}{\tikz{\node[draw,blue,scale=0.4,circle,line width = 1.0pt,fill=none](){};}}}
\newcommand{\gcircle}{\raisebox{0.5pt}{\tikz{\node[draw,black!60!green,scale=0.4,circle,line width = 1.0pt,fill=none](){};}}}

\begin{document}

\preprint{Physical Review Fluids}

\title{Coupled population balance and large eddy simulation model for polydisperse droplet evolution in a turbulent round jet”}

\author{Aditya Aiyer}
\email{aaiyer@princeton.edu}
\author{Charles Meneveau}
\affiliation{%
 Department of Mechanical Engineering, Johns Hopkins University, Baltimore, MD, 21218, USA\\
}%

\date{\today}
\begin{abstract}
 A population balance model coupled with large eddy simulations (LES) is adapted and applied to study the evolution of oil droplets in an axisymmetric turbulent jet including the effects of droplet breakup. A key unknown in simulating secondary breakup in turbulent multiphase jets is the inflow size distribution generated within the primary breakup zone near the nozzle exit. A mono-disperse injection inflow condition is commonly used for simplicity, but this choice is often unrealistic. In order to provide more realistic inlet conditions for LES, we develop a one dimensional (1D) parcel model to predict the evolution of the dispersed phase along the jet centerline due to the combined effects of advection, radial turbulent transport and droplet breakup due to turbulence in the regions closer to the jet nozzle that cannot be resolved using coarse LES. The model is validated with experimental data measured far from the nozzle. The 1D model is   also used to generate an initial size distribution for use in a coarse-resolution LES of a turbulent jet. Number density fields for each bin of the discretized droplet size distribution  are modeled using an Eulerian LES approach and scalar transport equations are solved for each bin.  LES results are compared to published experimental data, with good agreement and we examine the statistics of the velocity field and the concentration of the polydisperse oil droplet plumes for two droplet Weber numbers. We find that the centerline decay rate of the concentration for different droplet sizes is modified in the breakup dominated zone. Unlike Reynolds averaged approaches, LES also allows us to quantify size distribution variability due to turbulence. We quantify the radial and axial distributions and the variability of key quantities such as the Sauter mean diameter, total surface area and droplet breakup time-scale and explore their sensitivity to the Weber number. 

\end{abstract}

\maketitle

\section{Introduction}\label{sec:intro}

Accurate characterization of the  dispersed phase size distribution is crucial  in the context of numerous natural and engineering multiphase flow processes.  In liquid atomization, detailed information of the droplet size is of great importance  in the design and application of spray systems \cite{Movahednejad2010}. Of much interest during the Covid-19 epidemic, the size distribution of drops generated by coughing affects their residence time in the air as well as the ability of masks to prevent their transmission \cite{mittal_2020}. Conversely, for drug delivery \cite{Morita2016} one needs small drops that can be easily inhaled. In the context of underwater oil spills \cite{North2015}, the size distribution of oil droplets formed in the jet at the well strongly influences the fate and transport of oil in the aquatic environment \cite{Yang2014b}. Large droplets tend to rise more quickly to the surface, while the smaller ones may remain submerged for longer periods and are more dispersed due to turbulent mixing \cite{Johansen2003DevelopmentModels}. Typically, many important systems can be idealized to the case of droplet formation and breakup processes in turbulent jets and plumes.

There are numerous studies modelling droplet size distributions in turbulent systems due to the effects of breakup and coalescence. In the case of liquid atomization and spray systems, focus has been on modelling the characteristic droplet diameter. \citet{Lee2011CalculationEquations} developed a model for the mean diameter in a spray system using the integral form of the conservation equation.  In the case of oil droplet breakup, most previous studies focus on the steady state distribution in stirred tank reactors \cite{Calabrese1986,Prince1990,Tsouris1994}. There are few studies that model oil droplet breakup and formation in turbulent jets.  \citet{Bandara2011} coupled a population dynamics model with a plume model CDOG \cite{Zheng2003} to predict droplet size distributions from the DeepSpill experiments \cite{Johansen2003DeepSpill-FieldWater}. \citet{Johansen2013} established correlations for the characteristic diameter at steady state based on the Weber number and Reynolds number. They used analytical functions to fit the droplets size distribution at steady state. \citet{Zhao2014} developed a fluid parcel model VDROP-J to predict size distributions along the centerline of a turbulent jet by parameterizing the velocity and dissipation field of the jet. Their model neglected the effects of radial turbulent transport of the centerline concentration and was therefore limited to predicting the relative droplet size distribution at the centerline but not the actual number density. 

Modeling developments and testing must be informed by experimental data. Experimental studies of breakup of liquid jets primarily focus on the distribution of droplets far downstream of the nozzle. For instance, \citet{Eastwood2004} injected droplets of varying density, viscosity and interfacial tension into a fully developed water jet and tracked particle size distributions using digital image processing techniques. \citet{Brandvik2013} performed a series of oil jet experiments with varying injection conditions and measured the steady droplet size distributions. The concentration was measured using an in situ laser diffractometer. The apparatus had a maximum detection size of $460\;\mu\mbox{m}$ which could be insufficient for some of the cases simulated. Experimental data for multiphase jets in cross flow have also been collected \cite{Murphy2016} downstream of the nozzle. In all these experiments the droplet size distribution in the near nozzle region is difficult to characterize due to the high turbulence intensity and high opacity.  Recently there have been advances in using ultra-small angle x-ray scattering \cite{Kastengren2017} and refractive   index   matching  with  planar   laser-induced   fluorescence \citet{Xue2019} in order to probe the near nozzle region of the jet.

There has been significant progress in simulating two-phase flows
using high resolution grids that can describe detailed deforming interfaces and thus capture the formation of droplets from instabilities of liquid sheets \cite{Desjardins2008AnAtomization,Herrmann2010DetailedCrossflow,Tomar2010MultiscaleAtomization,Bravo2019EffectsRadiography}.  The aim is to, for example, simulate primary atomization and determine the resulting droplet size distributions. The focus of these simulations has been to determine the liquid core length and the resultant characteristic droplet diameter. 
However, in most applications, two-phase jet flows  are characterized by a large separation of scales ranging from a few hundred microns (droplet scale), to the order of millimeters (nozzle scale), and up to meters (jet far field or plume). In environmental applications this separation of scales can be even larger.  Direct numerical simulations of such systems becomes intractable due to the high computational cost involved in attempting to capture all relevant scales. Modeling approaches are needed, and focus on providing averaged or 'coarse-grained' solutions. Such approaches must make judicious choices weighing computational cost with accuracy and considering how much details about the simulated phenomena are needed. 
In the present work we consider levels of description typical of coarse-scale large eddy simulations (LES) and we follow the approach of Ref. \cite{Aiyer2019AEvolution} that couples population dynamics equations with Eulerian LES for both the continuous and dispersed phases (population densities of droplets of various sizes). 

\begin{figure}
        \centering
    
        \includegraphics[width=0.5\linewidth]{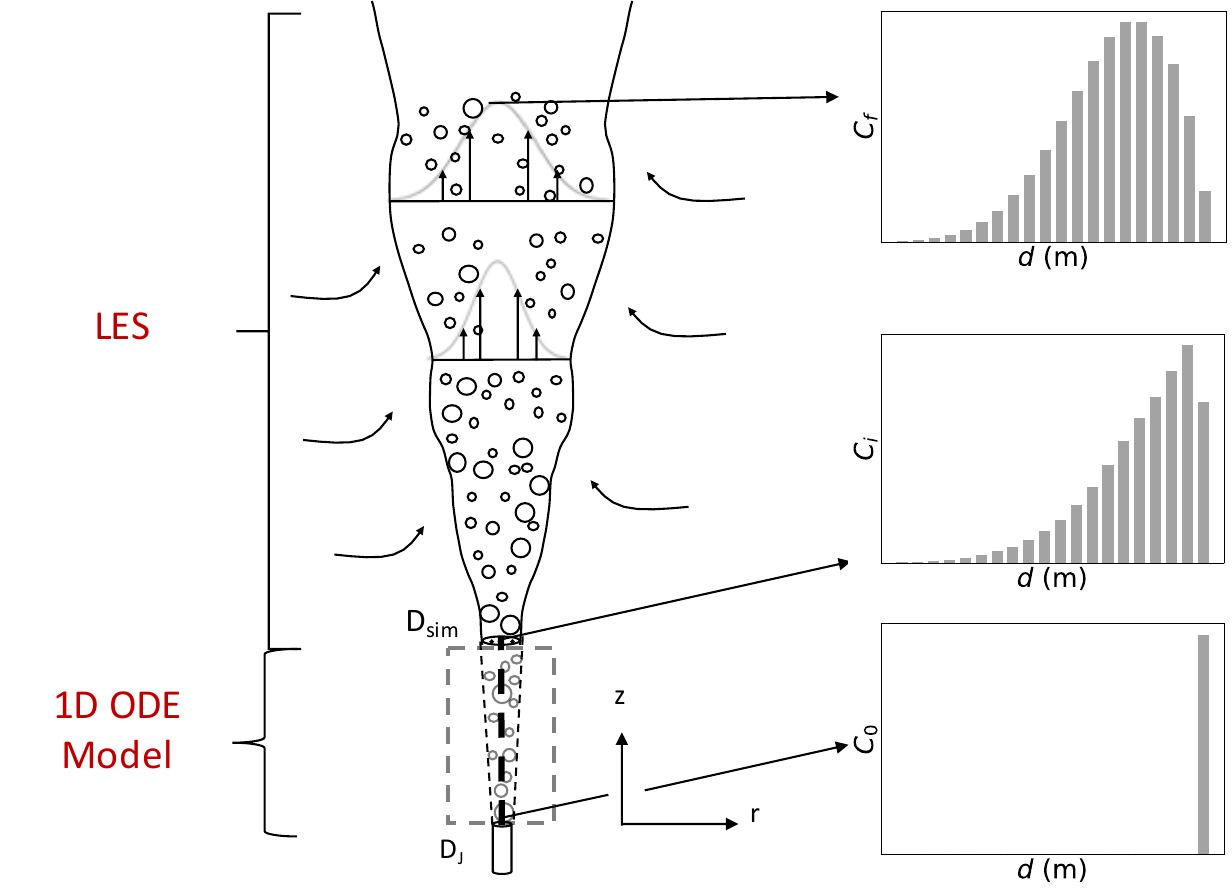}

  \caption{Schematic diagram of the hybrid ODE-LES modeling approach: Between the nozzle and the end of the 1D ODE model region, the size distribution is obtained by integrating a 1D ordinary differential equation for the centerline concentrations. The results are used as inflow concentrations for the Eulerian-Eulerian LES further downstream.}
    \label{fig:sketch_ODE}
\end{figure}

Even though limited to applications with relatively low volume fractions, Eulerian approaches can be advantageous since they are not constrained by the number of droplets, as the distribution of droplets in each size range is described by a continuous concentration field \cite{Pedel2014,Yang2016}. The approach used in Ref. \cite{Aiyer2019AEvolution}  focused on the secondary breakup and transport of droplets in a turbulent jet subject to a uniform crossflow. The model was validated by comparisons of the relative size distribution from their coarse mesh LES with laboratory experimental data \cite{Murphy2016}, showing reasonably good agreement. Such coarse LES have a significantly lower computational cost than DNS but do not resolve phenomena associated with the  primary breakup zone near the nozzle and the resulting near-nozzle size distributions. Instead, a mono-disperse size distribution is commonly used as inflow in such cases, but this is not sufficiently realistic. In the present paper, we develop a hybrid modeling approach that allows us to specify more realistic inlet size distributions for use in a coarse LES of a turbulent jet. As shown schematically in  figure \ref{fig:sketch_ODE} the LES starts some distance downstream of the experimental nozzle, where the initial jet has spread sufficiently so that it can be resolved by the coarse LES mesh. We do not model the primary breakup processes via fine-grid LES or DNS, but rather model droplet breakup due to turbulence at smaller scales than what we can resolve initially in our simulation using a one-dimensional version of the population dynamics approach (denoted as 1D ODE model) as a reduced order parcel model for these processes. Moreover, using simplified (eddy-viscosity based) theory of turbulent jet evolution  to account for the radial turbulent transport of centerline concentrations, the approach predicts the actual rather than the relative concentration distribution at the centerline. The 1D ODE population dynamics model is validated with experimental data available downstream, and is then used to provide the inflow size distribution (inlet  condition) for the proposed hybrid ODE-LES approach, as shown in figure \ref{fig:sketch_sim}.

 The paper begins in \S \ref{sec:ode_model} presenting the formulation and validation of one-dimensional reduced turbulent breakup model used to generate an inlet condition for LES. We present a LES of a turbulent jet with oil droplets injected at the centerline in \S \ref{sec:LES} and present results for two droplet Weber numbers in \S \ref{sec:results}. Conclusions are presented in \S \ref{sec:conclusions}.

\section{\label{sec:ode_model}one-dimensional ODE model}

Following the approach of Ref. \cite{Aiyer2019AEvolution}, the size distribution of drops is assumed to be governed by a population dynamics equation including the effects of advection, radial diffusion and droplet breakup due to turbulence. 
 In general, one can include other source terms for coalescence, evaporation or aggregation into the framework but here we focus on dilute turbulent jets and limit the discussion to droplet breakup. 
 
 \subsection{Model development}
 
 We begin using a 2D polar co-ordinate system to develop the model with $z$ as the axial and $r$ the radial coordinate. The origin is at $z = 0$ corresponding to the nozzle exit shown in figure \ref{fig:sketch_ODE}. We use a Reynolds-averaged Navier-Stokes (RANS) formulation (unlike the eddy-resolving LES case treated in Ref. \cite{Aiyer2019AEvolution}) coupled with a simple eddy-viscosity approach. The 2D approach is then cast as a 1D problem by only considering the centerline evolution ($r=0$). 
 
 The total concentration field of oil droplets is discretized into a finite number of bins based on the droplet diameter. The total concentration is related to the concentration in each bin  through the equation
\begin{equation}
    \overline{c}(z,r) = \sum_{i=1}^N \, V_i  \overline{n}_i (z,r),
\end{equation}
where, $N$ is the total number of bins used to discretize the droplet size range, $n_i$ is the number of droplets of size $d_i$ per $m^3$ of fluid and $V_i = \left(\pi/6\right)d_i^3 $ is the volume of a droplet of diameter $d_i$. The overbar denotes RANS averaging.  The population dynamics equation for the droplet concentration including the effects of advection, radial diffusion and droplet breakup can be written for each droplet size as,
\begin{equation}\label{eqn:jet_evol_eqn}
     \bw \frac{\partial \overline{n_i}V_i}{\partial z} + \bv\frac{\partial \overline{n_i}V_i}{\partial r} = S_{b,i} \, V_i  \, + \,\frac{D_T}{r} \frac{\partial}{\partial r} \left(r\frac{\partial}{\partial r} \overline{n_i}V_i\right),
\end{equation}
where  $\bv,\bw$ are the mean radial and axial velocities, respectively, and $S_{b,i} \, V_i$ is the droplet breakup source term to model the change of the concentration due to droplet breakup, to be described later.  The eddy diffusivity $D_T$ is assumed to be independent of radial position and only depend on $z$. The rise velocity of the individual droplets has been neglected as it would be small compared to the jet centerline velocity in the near nozzle region. The molecular diffusivity is also neglected since typically $D<<D_T$. 
 
 The mean velocity is modeled following the classic eddy viscosity approach. 
  The conservation of mass and momentum in a round turbulent jet, expressed in polar co-ordinates using the boundary-layer approximation, read:
\begin{equation}
 \frac{\partial \bw}{\partial z} + \frac{1}{r}\frac{\partial(r \bv)}{\partial r} =0 ,
\end{equation}

\begin{equation}
    \bw\frac{\partial \bw}{\partial z} + \bv\frac{\partial \bw}{\partial r} = 
     \nu_T \frac{1}{r}\frac{\partial}{\partial r} \left(r \frac{\partial \bw}{\partial r}\right)
\end{equation}
Above, $\nu_T$ is the $z$-dependent eddy viscosity. Again, the molecular viscous diffusion term is neglected in the high Reynolds number cases considered.

The  mean velocity profile  using the assumption of a similarity variable can be written as \cite{Pope2001},
\begin{equation}\label{eqn:sim_sol}
    \bw(z,r) = \bw_0(z) {f}(\eta),
\end{equation}
where $\eta = r/(z-z_0)$ is the similarity variable. For the case of $r-$independent eddy viscosity,  the resulting self-similar velocity profile $f(\eta)$  reads \cite{Pope2001,Law2006},
\begin{equation}\label{eqn:sim_vel}
 f(\eta) = \frac{1}{\left(1+\alpha^2\eta^2\right)^{2}},
\end{equation}
where the coefficient $\alpha$ is related to the spreading rate $S$ of the jet, $\alpha^2 = (\sqrt{2}-1)/S^2$.  
The downstream centerline velocity, jet width, and dissipation can thus be deduced:
\begin{equation}\label{eqn:dissip_east}
   \frac{\overline{w}_0}{\bw(z)} = \frac{1}{C_u} \left(\frac{z}{D_J} - \frac{z_0}{D_J}\right), \,\,\,\,\,\,\,\,\, r_{1/2} = S (z-z_0), \,\,\,\,\,\,\,\,\, \frac{\epsilon D_J}{\overline{w}_0^3} = C \left(\frac{z}{D_J} - \frac{z_0}{D_J}\right)^{-4},
\end{equation}
where $C_u=6$, $S=0.1$ and $C=65$ \cite{MARTINEZ-BAZAN1999,Su1999PlanarFlows,Hussein1994}  are empirically determined constants, $z_0$ is the virtual origin of the jet, and $D_J$ is the nozzle diameter.

Next, we consider the droplet concentration equation (\ref{eqn:jet_evol_eqn}). Similar to equation (\ref{eqn:dissip_east}), we aim for a formulation that describes the centerline concentration evolution as function of $z$ only, and must therefore replace the radial derivatives term with a suitable approximation. To this end we assume that that the relative radial dependence of the solution is unaffected by the source term. 
Setting $S_{b,i} = 0$ in equation \ref{eqn:jet_evol_eqn}, the concentration in each bin obeys 

the same evolution equation of the total mean concentration, $\overline{c}$, given by:
\begin{equation}\label{eqn:scalar_jet}
 \bw \frac{\partial \overline{c}}{\partial z} + \bv\frac{\partial \overline{c}}{\partial r} = \frac{D_T}{r} \frac{\partial}{\partial r} \left(r\frac{\partial}{\partial r} \overline{c}\right)
\end{equation}
that is amenable to solution using a similarity variable. 
To complete the similarity solution, one must express it in terms of the 
total scalar flux injected at the source, $Q_0$, defined as 

\begin{equation}\label{eqn:cons_flux}
     Q_0 = 2\pi \int_0^{\infty}  \bw(z,r) \, \overline{c}(z,r) \, r \, dr,
\end{equation}
which remains constant independent of $z$.

We can introduce a non dimensional scalar profile similar to equation (\ref{eqn:sim_sol}) for equation  (\ref{eqn:scalar_jet}) according to:
\begin{equation}\label{eqn:sim_conc}
    \frac{\overline{c_i}(z,r)}{\overline{c}_{0,i}(z)}\  = \ \phi(\eta)\  =\  \frac{\overline{c}(z,r)}{\overline{c}_{0}(z)} .
\end{equation}
In order to find $\phi(\eta)$ we can substitute equation (\ref{eqn:sim_conc}) into equation (\ref{eqn:scalar_jet}). Using the expressions for the self similar velocity profile and the evolution of the mean centerline velocity field described in equations (\ref{eqn:sim_vel}) and (\ref{eqn:dissip_east}), we obtain an ordinary differential equation (ODE) for $\phi(\eta)$, 
\begin{equation}\label{eqn:phi_int}
    D^{-1}_T\eta\left(1+\alpha^2\eta^2\right)^2\phi^{''} + \left(D^{-1}_T\left(1+\alpha^2\eta^2\right)^2 +\frac{C_uw_0 D_J\eta^2}{2}\left(1+\alpha^2\eta^2\right)\right)\phi^{'} + C_uw_0D_J\eta\phi =0.
\end{equation}{}
In order to write equation (\ref{eqn:phi_int}) in terms of $\alpha$ and the turbulent Schmidt number, $Sc_T = \nu_T/D_T$, we note that
for the constant eddy-viscosity solution, $C_uD_Jw_0 = 8\alpha^2\nu_T$ \cite{Pope2001}. Equation (\ref{eqn:phi_int}) can then be re-written as,
\begin{equation}
    Sc^{-1}_T\eta\left(1+\alpha^2\eta^2\right)^2\phi^{''} + \left[Sc^{-1}_T\left(1+\alpha^2\eta^2\right)^2 +4\alpha^2\eta^2\left(1+\alpha^2\eta^2\right)\right]\phi^{'} + 8\alpha^2\eta\phi =0.
\end{equation}

The solution to the above equation that monotonically decreases away from the centerline is given by \cite{Law2006},
\begin{equation}\label{eqn:phi_conc}
    \phi(\eta) = \frac{1}{\left(1+\alpha^2\eta^2\right)^{2 Sc_T}}.
\end{equation}{}
Equation \ref{eqn:phi_conc} is an exact solution to \ref{eqn:scalar_jet} but only approximately valid for the individual bin concentration fields as we had neglected the breakup source term in its derivation (for which no similarity solution exists in general). Note however that we only make this approximation in evaluating the radial derivative term, and then set $\eta=0$.
Substituting equations (\ref{eqn:sim_vel}) and (\ref{eqn:sim_conc})   into equation (\ref{eqn:jet_evol_eqn}) we obtain,
\begin{equation}\label{eqn:inter_conc}
      \bw_0(z)f(\eta)\frac{\partial}{\partial z}[ n_{0,i}(z)\phi(\eta) ] = S_{b,i} + 
       \frac{1}{r} \frac{\partial}{\partial r}\left(r {D_T} \frac{\partial }{\partial r} [n_{0,i}(z)\phi(\eta)] \right).
\end{equation}
Substituting the similarity solution given by equation (\ref{eqn:phi_conc}), evaluating the derivatives with respect to $r$ and setting $r=\eta=0$ we obtain the centerline evolution of each bin's number concentration, 
\begin{equation}\label{eqn:centerline_conc}
    \frac{d}{dz}{n_{0,i}(z)} = S_{b,i}(z,0) \, \frac{z}{w_0 C_u D_J} -  \frac{n_{0,i}(z)}{z} .
\end{equation}{}
Equation (\ref{eqn:centerline_conc}) describes a system of ODEs that needs to be solved numerically to obtain the evolution of the individual droplet concentrations, accounting for droplet breakup and turbulent transport at the centerline.
Note that the breakup source term $S_{b,i}$, does not alter the decay rate of the overall concentration, $\overline{c}(z)$ defined through equation (\ref{eqn:cons_flux}). This can be verified by multiplying equation (\ref{eqn:centerline_conc}) by the corresponding droplet volume $V_i$ and summing over all droplet sizes and noting that $\sum_i S_{b,i}V_i =0$. 

To complete the model description, we   summarize how the droplet breakup source term $S_{b,i}$ is modeled.  Following \cite{Aiyer2019AEvolution} we write:
\begin{equation}\label{eqn:break_source}
    S_{b,i}(z,0) = \sum_{j>i}^{N}P(d_i,d_j)g(z,0,d_j)\overline{n}_j(z,0,d_j) - g(z,0,d_i)\overline{n}_i(z,0,d_i).
\end{equation}{}
The first term on the right-hand side of equation $(\ref{eqn:break_source})$ represents the birth of droplets of size $d_i$ due to the total contribution from breakup events of larger droplets of 
diameter $d_j$. The second term accounts for death of droplets of size $d_i$ due to breakup.
$P(d_i,d_j)$ is the probability of formation of a droplet 
of size $d_i$ due to the breakup of a parent droplet of size $d_j$. The breakup is considered to be binary, and $P(d_i,d_j)$ is formulated based on the formation energy required to form the daughter droplets of size $d_i$ and a complementary droplet to ensure volume conservation \cite{Tsouris1994,Aiyer2019AEvolution}. 

The breakup frequency $g(z,0,d_i)$ is formulated based on the popular method of modelling breakup based on encounter rates of turbulent eddies and their characteristic fluctuations with droplets of a certain size \cite{Tsouris1994,Luo1996}. These models were limited to droplet-eddy collisions in the inertial range of turbulence. \citet{Aiyer2019AEvolution} extended these models, by using a second order structure function to characterize the eddy fluctuation velocity including the viscous range. 
The breakup frequency is expressed as a function of the Reynolds number ($Re_i$) based on droplet diameter and a velocity scale defined as $u(d_i)=(\epsilon d_i)^{1/3}$, the Ohnesorge number ($Oh_i$) of the dispersed phase controlling the relative importance of viscosity to surface tension of the droplet, and the density and viscosity ratio ($\Gamma_i$) of droplet to carrier flow fluid. These non-dimensional numbers are defined below :
\begin{equation}\label{eqn:nond}
Re_i = \frac{\epsilon^{1/3} d_i^{4/3}}{\nu}\quad; \quad Oh_i = \frac{\mu_d}{\sqrt{\rho_d \sigma d_i}}\quad;\quad
      \Gamma = \frac{\mu_d}{\mu_c}\left(\frac{\rho_c}{\rho_d}\right)^{1/2}
\end{equation}
The breakup frequency for a given value of $\Gamma = 5.45$ then takes the form,
\begin{equation}\label{eqn:br_final}
\begin{gathered}
g_i(Re_i,Oh_i;\Gamma) = \frac{ K^{*}} {\tau_{b,i}}\ 10^{G(Re_i,Oh_i)}, \\ G(Re_i,Oh_i) = a \, [\log_{10}(Re_i)]^b + c \, [\log_{10}(Re_i)]^d  - e ,
\end{gathered}
\end{equation}
where, $K^*=0.2$ is a  empirically determined \cite{Aiyer2019AEvolution} dimensionless constant and $\tau_{b,i}= \epsilon^{-1/3}d^{2/3}$ is the breakup timescale for an eddy of size equal to that of the droplet. The fits for parameters $a$--$e$ as functions of $Oh$ are provided in Ref. \cite{Aiyer2019AEvolution}.

\subsection{Model Validation}\label{subsec:validation}

 We validate the 1D ODE model in equation (\ref{eqn:centerline_conc}) with data from turbulent oil jet experiments performed by Ref. \cite{Brandvik2013}. The experimental setup consists of a cylindrical tank with a diameter of $3\;\mbox{m}$ and a height of $6\;\mbox{m}$ with crude oil being injected at a controlled temperature at various flow rates. The details of the cases used in this work are provided in table \ref{tab:expt_conditions}. The droplet size distribution for each case was measured $2\;\mbox{m}$ above the nozzle exit using a LIST-100X laser diffractometer. For each experiment oil volume fractions (i.e oil concentration of a particular bin, normalized by the total concentration of all bins) were provided for 29 logarithmically spaced droplet size classes ranging from $4.5\;\mu\mbox{m}$ to $460\;\mu\mbox{m}$. Droplets larger than $460\;\mu \mbox{m}$ could not be recorded by the instrument.   
 The experimental data is reported as a relative volume fraction of oil for each size at the measurement location. In order to make comparisons with the 1D ODE model, we first need to determine the total oil concentration of the reported distribution. 
 
 The overall centerline concentration can be determined as a function of downstream distance, Schmidt number, $\mbox{Sc}_T$ and the inflow rate $Q_0$ using equation (\ref{eqn:cons_flux}). Substituting the similarity profiles for $w(z,r)$ and $\phi(z,r)$ into the equation we obtain
\begin{equation}\label{eqn:int_flux}
   Q_0 = 2\pi z^2 \int_0^\infty \frac{w_0(z)}{\left(1+\alpha^2\eta^2\right)^2}\frac{c_0(z)}{\left(1+\alpha^2\eta^2\right)^{2Sc_T}} \,\eta\, d\eta.
\end{equation}{}
The integral in equation ($\ref{eqn:int_flux}$) can be evaluated and yields
$ c_0(z) w_0(z) (2 \alpha^2) ({2\mbox{Sc}_T +1})^{-1}$.
We can therefore evaluate the centerline concentration as a function of downstream distance $z$ and the centerline mean velocity $w_0(z)$ as:
\begin{equation}\label{eqn:downstream_conc}
    c_0(z) = \frac{Q_0\alpha^2(2\mbox{Sc}_T+1)}{\pi w_0(z) \, z^2}.
\end{equation}{}
Given the known total oil injection rate $Q_0$ in the experiments, the total (all sizes) centerline oil concentration as a function of downstream distance, $c_0(z)$,  can be obtained using equation (\ref{eqn:downstream_conc}). This result would be expected to include both the concentration measured by the instrument and the unmeasured concentration of larger drops.
 The limitation on the maximum measurable drop size is expected to lead to an underestimation of the total oil volume at the measurement location, since we expect at least some of the drops to be larger than $460\;\mu m$. An extrapolation approach will be used to augment the measurement data.

\begin{figure}
        \centering

        \subfloat[\label{fig:expt_3_5}]{\includegraphics[width=0.5\columnwidth,trim=4 4 4 4,clip]{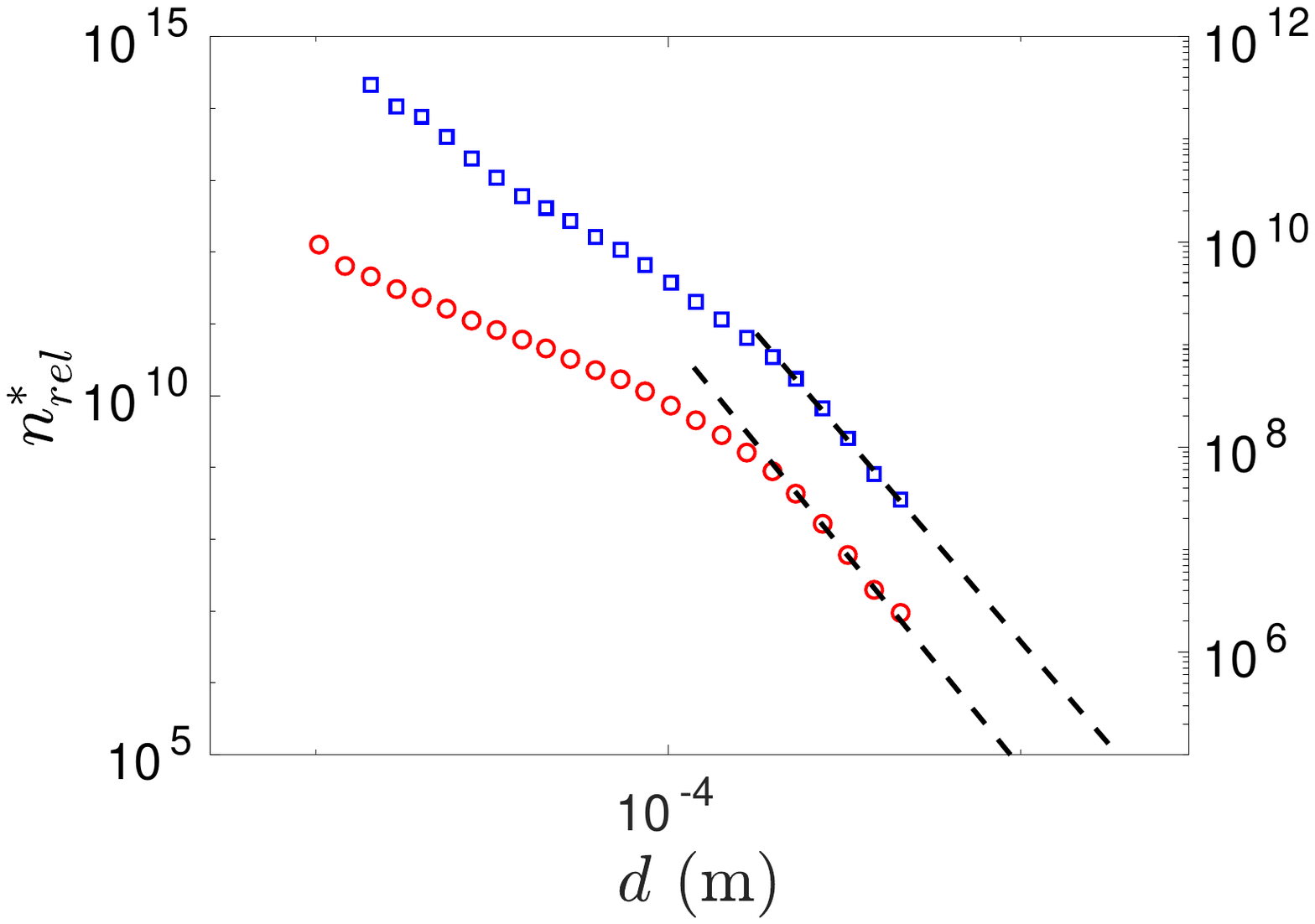}}
        \hfill
       \subfloat[\label{fig:dsd_3_5}] {\includegraphics[width=0.5\columnwidth,trim=4 4 4 4,clip]{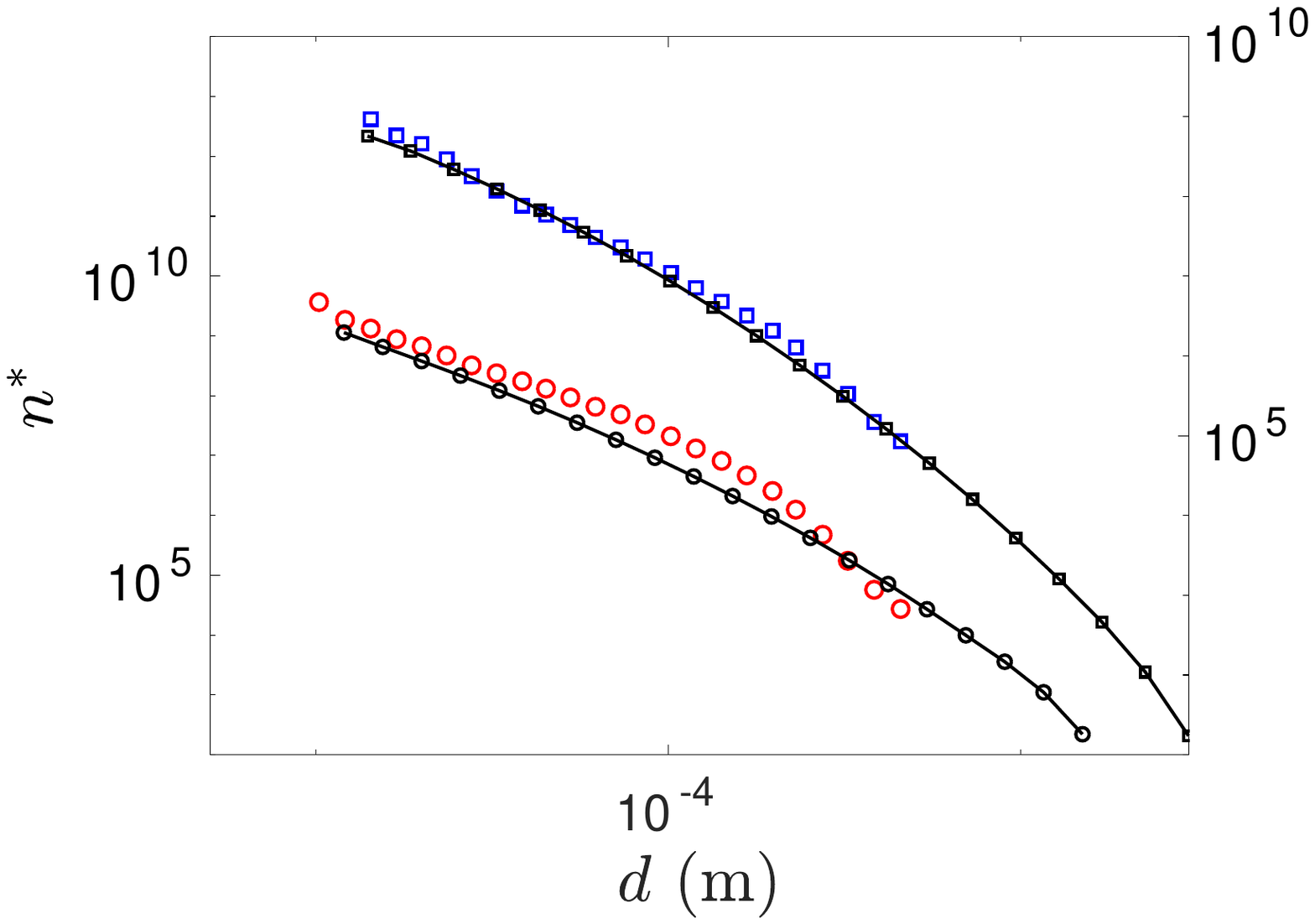}}
  \caption{ \protect\subref{fig:expt_3_5} Relative number density distribution of Expt. 1 (\protect\bsq) and Expt. 2 (\protect\rcircle) at $z=2m$.  The y-axis is scaled differently for visualization purposes. The right axis depicts the size distribution from Expt. 1 while the left axis depicts the size distribution from Expt. 2. The dashed line (\protect\blackdline) denotes the fit of the tail of the distribution with $F_1(d) = A_2 d^{-4}$ and $F_2(d) =A_1 d^{-6}$. The fitted constants are $A_1=1.28\times10^{-6}$ and $A_2=6.76\times10^{-14}$, \protect\subref{fig:dsd_3_5} Comparison of number density distribution from the 1D ODE model for Expt 1. (\protect\sqline) and Expt. 2 (\protect\circleline)  and corresponding experimental data (\protect\bsq , \protect\rcircle) at measurement location.} 
    \label{fig:num_3_5}
\end{figure}

 We define the number density, as the number concentration $\overline{n}_i$ normalized by the bin width, as the width of the bins used in the experiments is not necessarily the same as that used in the model, i.e 
\begin{equation}\label{eqn:num_dens}
    n^*_i = \frac{\overline{n_i}}{\delta d_i},
\end{equation}{}
where $\overline{n}_i$ is the number of droplets per $m^3$ fluid in bin $i$ with a bin width $\delta d_i = (d_{i+1}-d_{i-1})/2$ for $i=2$ to $N-1$, $\delta d_1 = d_2-d_1$ and $\delta d_{N} = d_{N}-d_{N-1}$. The normalization ensures that the result is independent of the discretization of the size range (bin width).
  The symbols in figure \ref{fig:expt_3_5} show   the experimentally measured relative number density $n^*_{rel}$ as a function of drop diameter at the measurement location $z=2\;m$, for both Expt. 1 and Expt. 2.
  The units for the relative number density are number of droplets per $m^3$ of fluid per bin width $\mu m^{-1}$ normalized by the total oil concentration of the measured distribution.
  We can  see that the scaling of the relative size distribution follows two distinct power law regimes for the small and large droplets. In order to quantify the unmeasured concentration, we smoothly extend the tail of the distribution to the nozzle diameter $D_J$ using a fitting function $F(d)$. This step will account for the contribution of droplets of size $d>d_{max}$ to the total concentration. The unmeasured volume fraction can then be calculated by integrating the fitted particle size distribution $F(d)$ from $d_{max}=460\;\mu\mbox{m}$ to the largest possible droplet size, here assumed to be the nozzle diameter $D_J$,
 \begin{equation}\label{eqn:unm_vf}
    \phi_{un} = \int_{d_{max}}^{D_{j}} v(d) \, F(d) \, \mathrm{d}d, 
\end{equation}
where $v(d) = \pi/6\ d^3$ is the volume of the particle with diameter $d$ (internal coordinate for the size range discretization). The concentration in the experimental distribution can be calculated as,
\begin{equation}\label{eqn:conc_dist}
    c_{dist} = \frac{c_0(z=2m)}{1+\phi_{un}}. 
\end{equation}
 
Using equations (\ref{eqn:downstream_conc}), (\ref{eqn:unm_vf}) and (\ref{eqn:conc_dist}) we can determine the fraction of the concentration measured by limiting the maximum diameter to $d_{max} = 460\;\mu\mbox{m}$. For the case with $d=1.5\;\mbox{mm}$ and $Q=1.5\;L/\mbox{min}$ we find that the measured concentration accounts for $92.86\%$ of the total, whereas for $d=3\;\mbox{mm}$ and $Q=5\;L/\mbox{min}$ the measured concentration describes only $44\%$ of the total concentration. Thus for the case with the larger nozzle diameter, restricting the maximum droplet size to $460\;\mu\mbox{m}$ would underestimate the total volume of oil measured. Therefore, for our validations and subsequent simulations, we choose the maximum droplet size to be equal to that of the nozzle.

For the purpose of validation, we discretize the droplet size range into $N = 20$ bins, with the maximum diameter, $d_{20} = D_J$. We have tested the sensitivity of the results to the number of bins used to discretize the droplet size range and find that $20$ bins is sufficient to accurately capture the size distribution.
The initial droplet concentration is determined by equation (\ref{eqn:downstream_conc}) at a distance of $z=2\;D_J$ from the nozzle and a potential core region is assumed between $2-6\;D_J$ after which the velocity and dissipation decay according to equation ($\ref{eqn:dissip_east}$). Experimental data [16,32] suggest that self-similarity is valid starting from $z \approx 15-20\ D_J$ for the mean axial velocity and  $z \approx 10\ D_J$ for the mean dissipation rate. As will be shown in the subsequent section on model validation, we find that assuming that the self-similar solution is valid even in the near nozzle (from $z = 6-10\ D_J$) leads to accurate model predictions for the far-field (see figure 2).

The entire inflow number concentration was placed into the largest bin, $d_{20}$, i.e. assuming that the injection begins at the nozzle with drops of diameter equal to the jet diameter. The number concentration was obtained by dividing the concentration in a bin by the volume of a single drop with diameter equal to that corresponding to the bin size. The concentration for all other bins were initially set to zero. The physical properties of the oil and inflow conditions of the experiments are summarized in Table \ref{tab:expt_conditions}. The Schmidt number is set to $Sc_T = 0.7$ which is in the range of commonly used values in literature for passive scalars in turbulent jets \cite{Chua1990,Lubbers2001}.
We then numerically solve the set of ODE's Eq. (\ref{eqn:centerline_conc}) for the number concentration (number of droplets per $m^3$ of fluid) of each droplet size, from $z = 2\;D_J$ to $z = 666\;D_J$ (corresponding to $z=2\;m$), also using equations (\ref{eqn:dissip_east}),(\ref{eqn:downstream_conc}), and (\ref{eqn:break_source}).

The total experimental distribution for each case was calculated by multiplying the relative size distribution by the total concentration obtained from equation (\ref{eqn:conc_dist}). This renormalization ensures that the total oil flux at the measurement location in the experiment is equal to the source flux $Q_0$.
The number density from the model is compared to the experimental results at $z=2\;\mbox{m}$ for Expt. 1 and Expt. 2 in figure \ref{fig:dsd_3_5}. 
We see that the model not only predicts the size distribution in the experimental size range, but also smoothly extends the distribution for larger sizes. The total concentration distribution can then be reconstructed using equation (\ref{eqn:sim_conc}).
\begin{table*}
\caption{\label{tab:expt_conditions}Summary of Experimental conditions for the different cases.}
\begin{ruledtabular}
\begin{tabular}{cccccc}
$No.$&  $D_J$ (mm)& $Q$ ($L/min$) & $U_J$ ($m/s$) & $\sigma (N\;m^{-1})$ & $\mu_d (Pa\;s)$\\[5pt] \hline
Expt. 1 &3 & 5 & $11.8$ & $15.5\times 10^{-3} $ & $5\times10^{-3}$ \\
Expt. 2& 1.5 & 1.5 & $14.1$ & $15.5\times 10^{-3} $ & $5\times10^{-3}$ \\
\end{tabular}
\end{ruledtabular}
\end{table*}

\subsection{Inlet condition for Large Eddy Simulations}\label{subsec: Inlet_dist}

In the previous section, it was shown that the 1D ODE model can predict the average size distribution of oil droplets at the centerline, showing good agreement with experimental datasets. This model can be considered as sufficient if the only aim is to predict the time-averaged size distribution. If one also wishes to predict the  variability of the size distribution and radial concentration fluctuations in each size bin, taking into account the effect of the underlying turbulence, the use of  LES is required \cite{Aiyer2019AEvolution}. 
In this section we describe using the 1D ODE model to generate an inlet condition for LES bridging the near nozzle region to further downstream, where LES begins to resolve the flow. We explain the approach for the case with $D_J = 3\;mm$ and $Q = 5\; L/min$ and assume that the LES grid is coarse such that only at $z>10 D_J$ can it begin to represent the eddying motions inherent in the turbulent jet. 

\begin{figure}
        \subfloat[\label{fig:vel_dissip_input}]{\includegraphics[width=0.5\columnwidth]{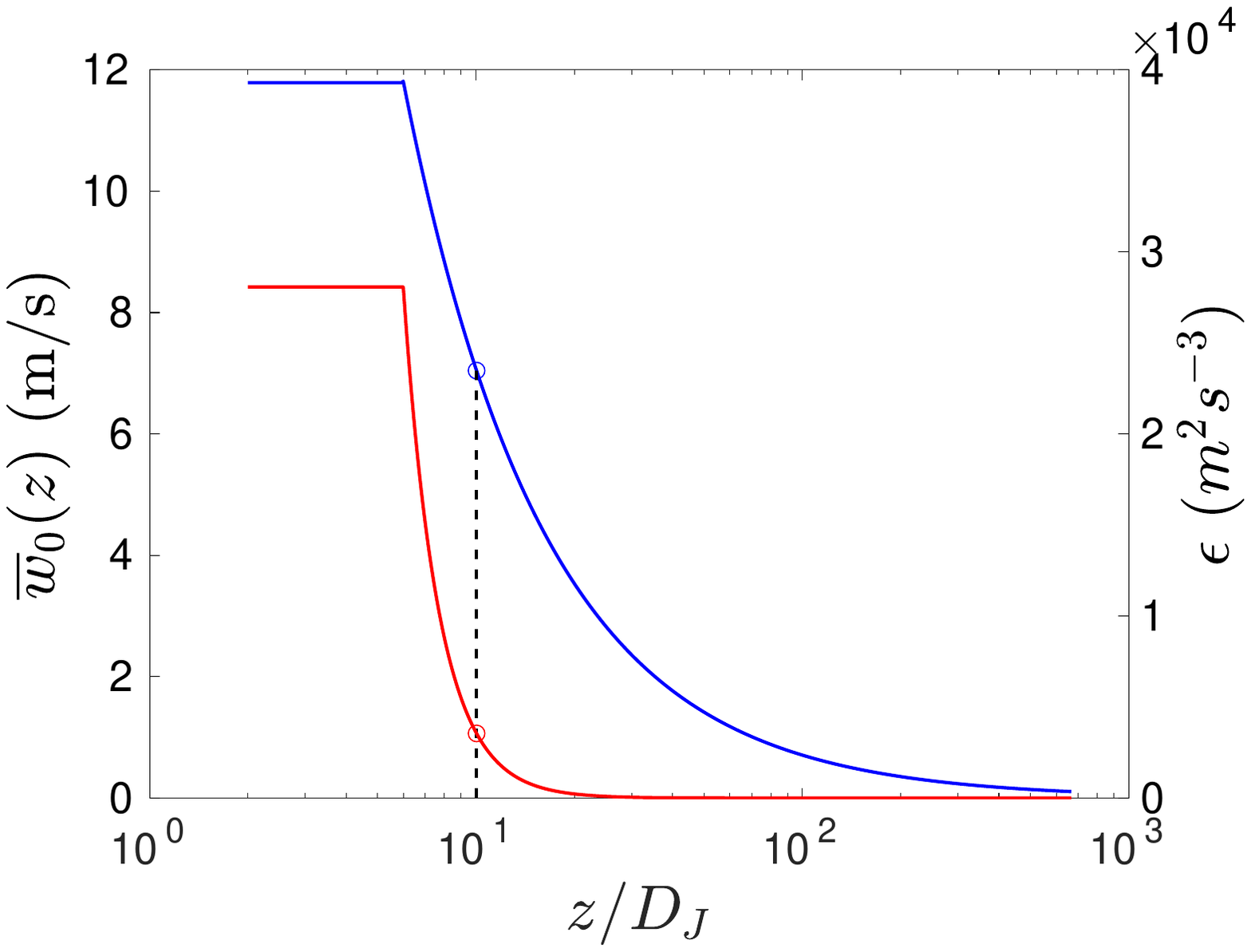}}
        \hfill
        \subfloat[\label{fig:conc_cent_output}
]{\includegraphics[width=0.5\linewidth]{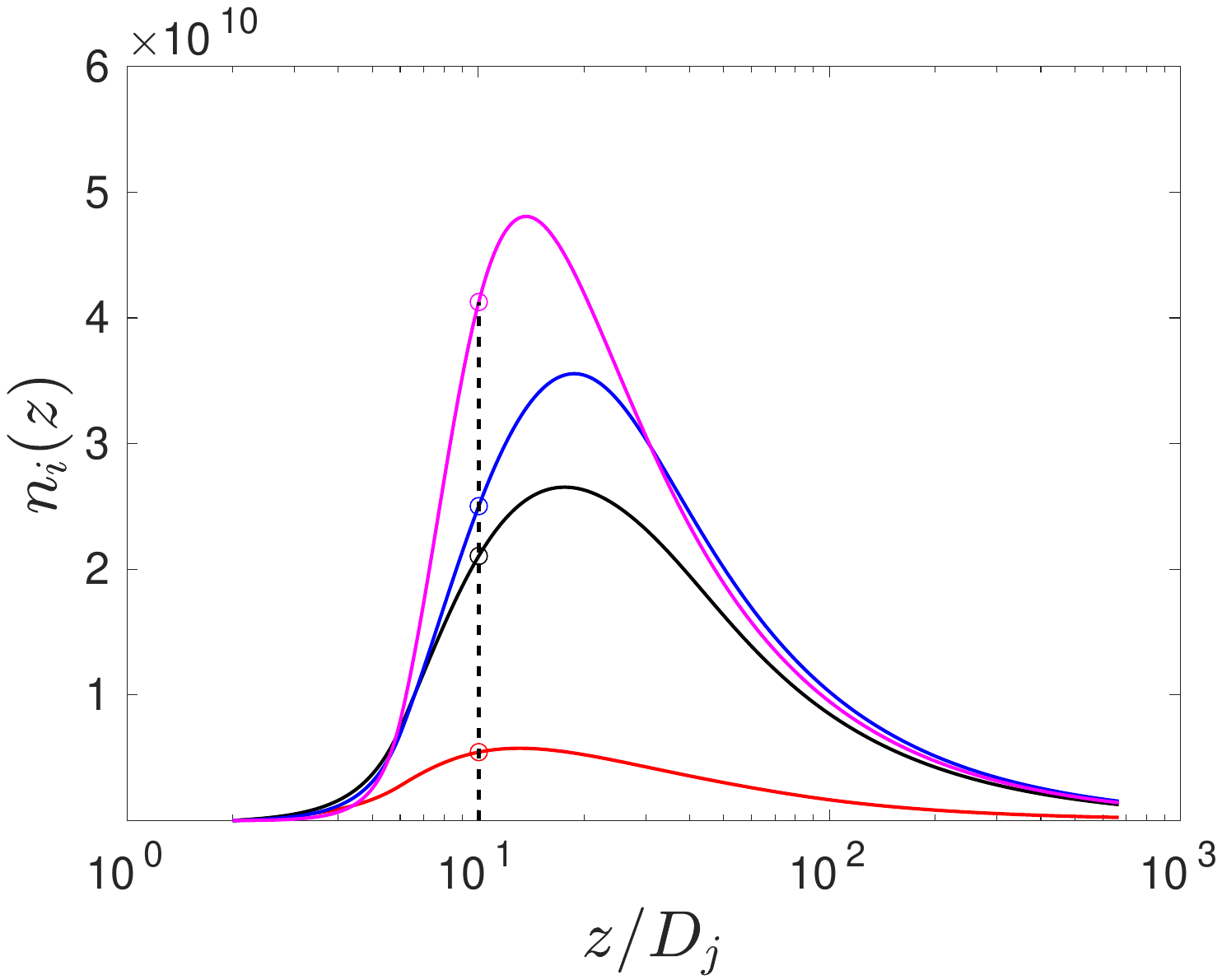}}
  \caption{ \protect\subref{fig:vel_dissip_input} Parameterized jet centerline velocity (\protect\blueline) and dissipation (\protect\redline), used as an
input to the 1D ODE model. \protect\subref{fig:conc_cent_output} Scaled centerline number concentration, $n_2$ ; $d = 18.5\;\mu \mbox{m}$ (\protect\magline), $5\times n_7$ ; $d = 76\;\mu \mbox{m}$ (\protect\blueline), $10\times n_9$ ; $d = 134\;\mu \mbox{m}$(\protect\blackline) and $10\times n_{12}$ ; $d =313\;\mu \mbox{m}$(\protect\redline) as a function of downstream distance. The initial conditions for LES are determined by the concentration values at $z=10 D_J$ depicted by the dashed line (\protect\blackdline).}
    \label{fig:ODE_inp_output}
\end{figure}

The 1D ODE model requires as input the centerline velocity and dissipation, for which we utilize Eq. (\ref{eqn:dissip_east}).  These inputs are plotted in figure \ref{fig:vel_dissip_input}, where we plot the evolution of the centerline velocity and dissipation as a function of downstream distance. The 1D ODE model (Eq. (\ref{eqn:centerline_conc})) is then integrated numerically and the results, i.e. the normalized centerline concentration evolution of the different droplets sizes, are shown in figure \ref{fig:conc_cent_output}. As can be seen, in the first part the breakup process dominates the evolution (concentration of the smaller drop sizes increases downstream), while downstream (after around $z/D_J \sim 30$), all concentrations decrease monotonically, where fluid transport (axial advection and radial turbulent transport) dominates the evolution of concentrations.
In order include some of the droplet breakup process in the LES domain, we  chose   $z = 10 \;D_J$ downstream of the nozzle exit as the location where the size distribution from the 1D ODE model is used as inlet condition for the LES. This location is depicted by the dashed line in figure \ref{fig:conc_cent_output}.  The jet width at this location can be calculated using equation (\ref{eqn:dissip_east}) to be $r_{1/2} = 0.1 z = D_J$ (this width also sets the diameter of an equivalent ``coarse jet for the LES'' as $D_{sim} \approx 2 D_J$, see discussion in \citet{Aiyer2019AEvolution}) . The corresponding centerline velocity shown in figure \ref{fig:vel_dissip_input} at $z = 10\;D_J$ is used as the jet injection velocity. It is important to note that there is no special significance of choosing $z=10\;D_J$. If a different location, for example, for $z=13\;D_J$, we would use the corresponding size distribution from figure \ref{fig:conc_cent_output} and injection velocity from figure \ref{fig:vel_dissip_input}. 

 \section{LES of Polydisperse droplets in a Turbulent Jet}\label{sec:LES}
 Large-eddy simulations (LES) are effective in resolving the large and intermediate scale structures of a turbulent flow, and only require modeling of the unresolved subgrid turbulent effects. The particles are modeled using an Eulerian description as concentration fields for each droplet size.
 This method has been successfully implemented to study mono-disperse plumes \cite{Yang2014,Yang2015,Chen2016,Yang2016} and polydisperse oil plumes in \cite{Aiyer2019AEvolution}.
 In the following sections, we review the numerical methods used for the LES, briefly describe the simulation setup. 
 \subsection{Eulerian-Eulerian LES equations}
 
Let $\xx = (x,y,z)$ with $x$ and $y$ be the horizontal coordinates and $z$ the vertical direction, and let $\ww = (u,v,w)$ be the corresponding velocity components. The jet and surrounding fluid are governed by the three--dimensional incompressible filtered Navier--Stokes equations with a Boussinesq approximation for buoyancy effects:
\begin{equation}\label{eqn:div}
\nabla \cdot \tuu =0,
\end{equation}
\begin{equation}\label{eqn:Navier_stokes}
\frac{\partial \tuu}{\partial t} + \tuu \cdot\nabla\tuu = -\frac{1}{\rho_c}\nabla\tp - \nabla\cdot \ttau^d +  \left(1 - \frac{\rho_d}{\rho_c}\right)\sum_i(V_{d,i}\tn_i)g\ee_3 +\tilde{F}\ee_3.
\end{equation}
\begin{equation}\label{eqn:conc}
\frac{\partial \tn_i}{\partial t} + \nabla \cdot (\tvv_i\tn_i) + \nabla \cdot \ppi_{i} = \widetilde{S}_{b,i} + \widetilde{q}_{i}, \,\,\,\, i=1,2.. N.
\end{equation}
A tilde denotes a variable resolved on the LES grid, $\tuu$ is the filtered fluid velocity, $\rho_d$ is the density of the 
droplet, $\rho_c$ is the carrier fluid density, $V_{d,i}=\pi d_i^3/6$ is the volume of a spherical droplet of diameter $d_i$, $\ttau = (\widetilde{\uu\uu} - \tuu\tuu)$ is the subgrid-scale stress tensor, $\tn_i$ is the resolved 
number density of the droplet of size $d_i$, $\widetilde{F}$ is a locally acting upward body force to simulate the jet momentum injection, and $\ee_3$ is the unit vector in the vertical direction. 
The filtered version of the transport equation for the number density $\widetilde{n}_i(\xx,t;d_i)$ is given by equation (\ref{eqn:conc}). The term $\ppi_{i} = (\widetilde{\vv_in_i} - \tvv_i\tn_i)$ is the subgrid-scale concentration flux of oil droplets of size $d_i$ (no summation over \textit{i} implied here) and $\widetilde{q}_i$ denotes the injection rate of droplets of diameter $d_i$.
In order to capture a range of sizes the number density is discretized into $N = 20$ logarithmically distributed bins for droplets between $d_1 =  14\;\mu\mbox{m}$ up to $d_{N_d} =  3\;\mbox{mm}$.  We then solve $N$ separate transport equations for the number densities $\widetilde{n}_i(\xx,t;d_i)$  with $i=1,2,...,20$.

Closure for the SGS stress tensor $\ttau^d$ is obtained from the Lilly-Smagorinsky eddy viscosity model with a Smagorinsky coefficient $c_s$ determined dynamically during the simulation using the Lagrangian averaging scale-dependent dynamic (LASD) SGS model \citep{Bou-Zeid}. The SGS scalar flux $\ppi_{i}$ is modelled using an eddy-diffusion SGS model. We use the approach of prescribing a turbulent Schmidt/Prandtl number, $Pr_{\tau} = Sc_{\tau} =0.4 $ \citep{Yang2016} (not to be confused with the RANS-level diffusivity and Schmidt number used in the previous section for the 1D model). The SGS flux can be parameterized as $\pi_{n,i} = -(\nu_{\tau}/Sc_{\tau})\nabla\tn_{i}$.
 With the evolution of  oil droplet concentrations being simulated, their effects on the fluid velocity field are modelled and implemented in  $(\ref{eqn:Navier_stokes})$ as a  buoyancy force term (the last term on the right-hand side of the equation) using the Boussinesq approximation. A basic assumption for treating the oil droplets as a Boussinesq active  scalar field being dispersed by the fluid motion is that the volume and mass fractions of the oil droplets are small within a computational grid cell.
   
The droplet transport velocity $\tvv_i$ is calculated by an expansion in the droplet time 
scale $\tau_{d,i}=(\rho_d+\rho_c/2)d_i^2/(18\mu_f)$ \citep{Ferry2001}. The 
expansion is valid when $\tau_{d,i}$ is much smaller than the resolved fluid time scales, 
which requires us to have a grid Stokes number $St_{\Delta,i} = \tau_{d,i}/\tau_{\Delta} \ll 1$, where $\tau_{\Delta}$ is the turbulent eddy turnover time at scale $\Delta$. 
The transport velocity of droplets of size $d_i$, $\tvv_i$,  is given by \citep{Ferry2001}
\begin{equation}\label{eqn:rise_vel}
\tvv_i = \tuu + w_{r,i}\ee_3 + (R-1)\tau_{d,i}\left(\frac{D\tuu}{D t} + \nabla \cdot \ttau\right) +O(\tau_{d,i}^{3/2}),
\end{equation}
where $w_{r,i}$ is the droplet terminal (rise) velocity, $\ee_3$ is the unit vector in the 
vertical direction, and $R = 3\rho_c/(2\rho_{d}+\rho_c)$ is the acceleration parameter. The droplet concentration transport velocity field has been modelled using the Fast Eulerian approximation which includes effects of drag, buoyancy, added mass and the divergence of the subgrid stress tensor. A more detailed discussion of the droplet rise velocity in equation $(\ref{eqn:rise_vel})$ can be found in \citet{Yang2016}.

 The term $\widetilde{S}_{b,i}$ in equation ($\ref{eqn:conc}$) represents the rate of change of droplet number density due to breakup.  The breakup rate $g_i$ is evaluated using the fits as in \cite{Aiyer2019AEvolution} that depend on the local Reynolds number expressed in terms of the local rate of dissipation. From the SGS model, the local rate of dissipation at the LES grid scale is given by
 \begin{equation}
 \epsilon({\xx},t) = 2 (c_s\Delta)^2|\tS| \tilde{S}_{ij}\tilde{S}_{ij}.
 \label{eq:dissipLES}
 \end{equation}

The equations ($\ref{eqn:div}$) and ($\ref{eqn:Navier_stokes}$) are discretized using a pseudo-spectral method on a collocated grid in the horizontal directions and a centered finite difference scheme on a staggered grid in the vertical direction \citep{Albertson1999}. Periodic boundary conditions are applied in the horizontal directions for the velocity and pressure field.
The transport equations for the droplet number densities, equation ($\ref{eqn:conc}$),
are discretized as in \citet{Chamecki2008}, by a finite-volume algorithm with a bounded third-order upwind scheme for the advection term.  A fractional-step method with a second-order Adams--Bashforth scheme is applied for the time integration, combined with a standard projection method to enforce the incompressibility constraint. 

\subsection{Simulation Setup}

\begin{figure}
        \centering
    
        \includegraphics[width=0.7\linewidth]{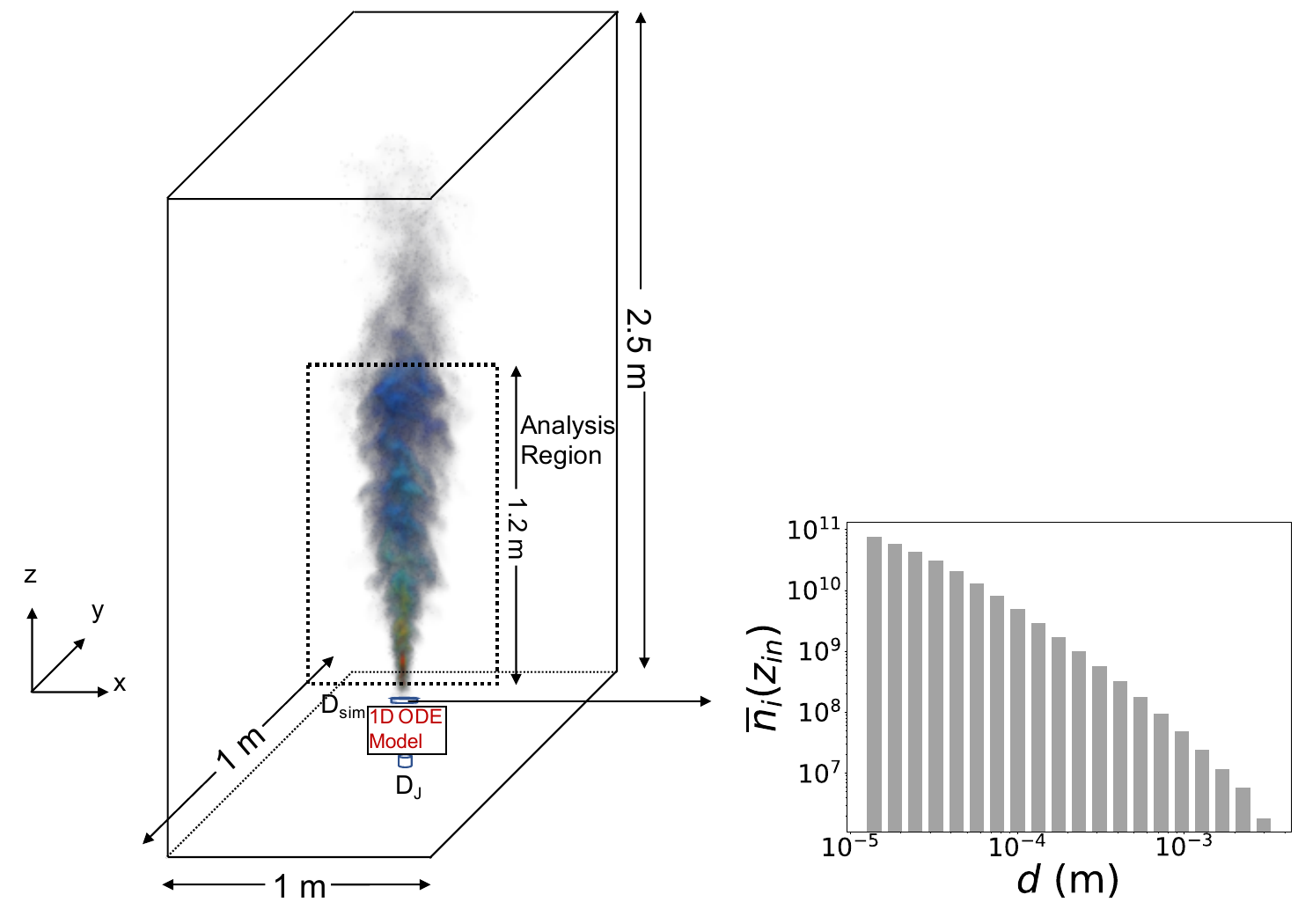}

  \caption{(a) Sketch of the simulation setup. Volume rendering of the instantaneous $14\;\mu\mbox{m}$ diameter droplet concentration with the $1000\;\mu\mbox{m}$ droplets visualized as dots placed randomly with density proportional to its concentration field. (b) Inlet distribution, $n_i$ (number of droplets per $m^3$ of fluid) for LES determined by the one-dimensional model,
}
    \label{fig:sketch_sim}
\end{figure}

\begin{table}
\centering
\begin{ruledtabular}

\begin{tabular}{c c c  c c c } 
 \centering No.
 & \multicolumn{1}{p{3cm}}{\centering 
 $We=\frac{2\rho_c\langle\epsilon\rangle^{2/3}d_{20}^{5/3}}{\sigma}$} & $\sigma (mN\;m^{-1})$ &$\overbar{w}_{in}$ (m/s) & $D_{sim}$ (mm) &$D_{J}$ (mm)\\ [5pt] \hline
 SIM 1 & 410 & 15.5 & 7 & 6  & 3\\ 
 SIM 2  & 820 & 7.75& 7 & 6  & 3\\ 
\end{tabular}
\caption{Simulation parameters.}
\label{tab:Sim_param}
\end{ruledtabular}{}
\end{table}

A sketch of the simulation domain is shown in figure $\ref{fig:sketch_sim}$. We simulate a turbulent jet aiming to reproduce the experiments of \citet{Brandvik2013}, specifically the case with nozzle diameter $D_{J}=3\;mm$ and flow rate $Q_0 = 5\;L/min$. This particular case, due to the larger nozzle diameter, allows us to use a relatively coarse mesh for the LES, while at the same time resolving a significant portion of the breakup. For instance, the case with $D_J=1.5\;mm$ discussed in section \S \ref{subsec:validation} would require us to have double the resolution in the horizontal directions in order to simulate the breakup dominated zone in LES.
The experimental setup and measurement techniques have been described in \ref{subsec:validation}. We use a hybrid approach where a population balance model is used to provide the drop concentration injection rates at each size ($q_{i}$) as inlet condition (figure \ref{fig:sketch_sim}b), and the subsequent secondary breakup and evolution of the oil droplets is simulated using LES. As shown in  figure $\ref{fig:sketch_sim}$, the simulations are carried in a rectangular box of size $(L_x,L_y,L_z) = (1,1,2.5)\;\mbox{m}$. The horizontal domain size has been chosen to be large enough so that boundary effects due to the periodic boundary conditions can be neglected. For instance, the total horizontal extent of the top of the analysis domain (at $z = 1.2m$) is $~ 10\ r_{1/2}$. As will be shown below, the axial velocity is negligible  for $r > 2.5\ r_{1/2}$. The experimental nozzle exit is chosen as the origin in the vertical direction. The simulated jet starts at a distance of $z= 10\;D_{J}$ from the origin.
 The simulations use a grid with $N_x \times N_y \times N_z = 288 \times 288 \times 384$ points for spatial discretization, 
and a timestep $\Delta t=6\times 10^{-5}\;\mbox{s}$ for time integration. The resolution in the horizontal directions,  $\Delta x = \Delta y = 3.47\;\mbox{mm}$ is set to ensure that at the location where the LES begins to resolve the jet (the `simulated inlet', see below), we have at least 3 points across the jet. In the vertical direction we use a grid spacing of $\Delta z = 6.5\;\mbox{mm}$ enabling us to capture a domain height 2.5 times the horizontal domain size.  
 
The injected jet is modelled in the LES using a locally applied vertically upward pointing body force following the procedure outlined in \citet{Aiyer2019AEvolution}, since at the LES resolution used in the simulation it is not possible to resolve the small-scale features of the injection nozzle. 
Random fluctuations are added to the horizontal components of the momentum equation to induce transition to turbulence. The fluctuations have an amplitude with a root-mean-square value equal to $0.1$ \% the magnitude of the forcing $\tilde{F}$ and are applied only during an initial period of $0.5\ s$ at the forcing source. The forcing is only applied over a finite volume and smoothed using a super-Gaussian kernel.
The method has been validated and presented in greater detail in the Appendix of \citet{Aiyer2019AEvolution}.
The resulting injection velocity is controlled by the strength of the imposed body force $\widetilde{F}$ applied such that the resulting centerline velocity in the LES matches the mean centerline velocity expected for the experiment at a distance $z= 10\;D_J$ from the experimental nozzle as shown in figure \ref{fig:vel_dissip_input}.
The droplet number density fields are initialized to zero everywhere. In order to avoid additional transient effects, the concentration equations are solved only after a time at which the jet in the velocity field has reached near the top boundary to allow the flow to be established. Based on the inlet distribution calculated in section \ref{sec:ode_model} oil droplets are injected as follows:   The number density transport contains a source term, $\widetilde{q}_i$ on the RHS of equation (\ref{eqn:conc}) 
that represents injection of droplets of a particular size.  The source term is calculated based on equation (\ref{eqn:int_flux}) for each bin size as:
\begin{equation}
    \widetilde{q}_{i} =\frac{\gamma_z\gamma_{xy}}{\Delta x\Delta y\Delta z}\frac{ 4\pi\alpha^2 z_{in}^2\overline{w}_0(z_{in})\overline{n_i}(z_{in}) (2\alpha^2)}{(2\mbox{Sc}_T+1)},
\end{equation}
where $\overline{w}_0(z_{in})$, $\overline{n_i}(z_{in})$ are the inlet velocity and concentration determined in \S \ref{subsec: Inlet_dist} at $z_{in} = 10\;D_J$. 
This ensures that the total injected concentration flux at the inlet $\sum_i \widetilde{q}_iV_i$ is equal to the source flux, $Q_0 = 5\;L/\mbox{min}$ from \citet{Brandvik2013}. The source is centered at ($x_c$, $y_c$, $z_c$) = (0.5 m, 0.5 m, 10 $D_J$) and distributed over two grid points in the z direction with weights $\gamma_{z} = 0.7$  and $\gamma_{z}=0.3$ at $z_c$ and $z_{c}+\Delta z$ respectively and over three grid points in the horizontal directions with weights $\gamma_{xy} = 0.292$ at $(x_c,y_c)$ and $\gamma_{xy} = 0.177$ at $(x_{c}\pm \Delta x,y_{c}\pm \Delta y)$.

In order to study the effects of changing Weber number on the concentration distribution, we perform a second simulation halving the surface tension of the oil, and thus doubling the Weber number.
The physical properties of the oil and the simulation parameters are given in Tables $\ref{tab:expt_conditions}$ and $\ref{tab:Sim_param}$.
 \section{Results}\label{sec:results}

 \subsection{Jet velocity and total concentration field}

 Statistics of the velocity and concentration fields are shown using a cylindrical coordinate system with $z$ being the axial coordinate, and supplement the time averaging with additional averaging over the angular $\theta$ direction. The LES averaged quantities will be denoted by angluar brackets while the averaged quantities from the 1D ODE model will be denoted by an overbar.

\begin{figure}
        \centering
        \includegraphics[width=0.6\linewidth]{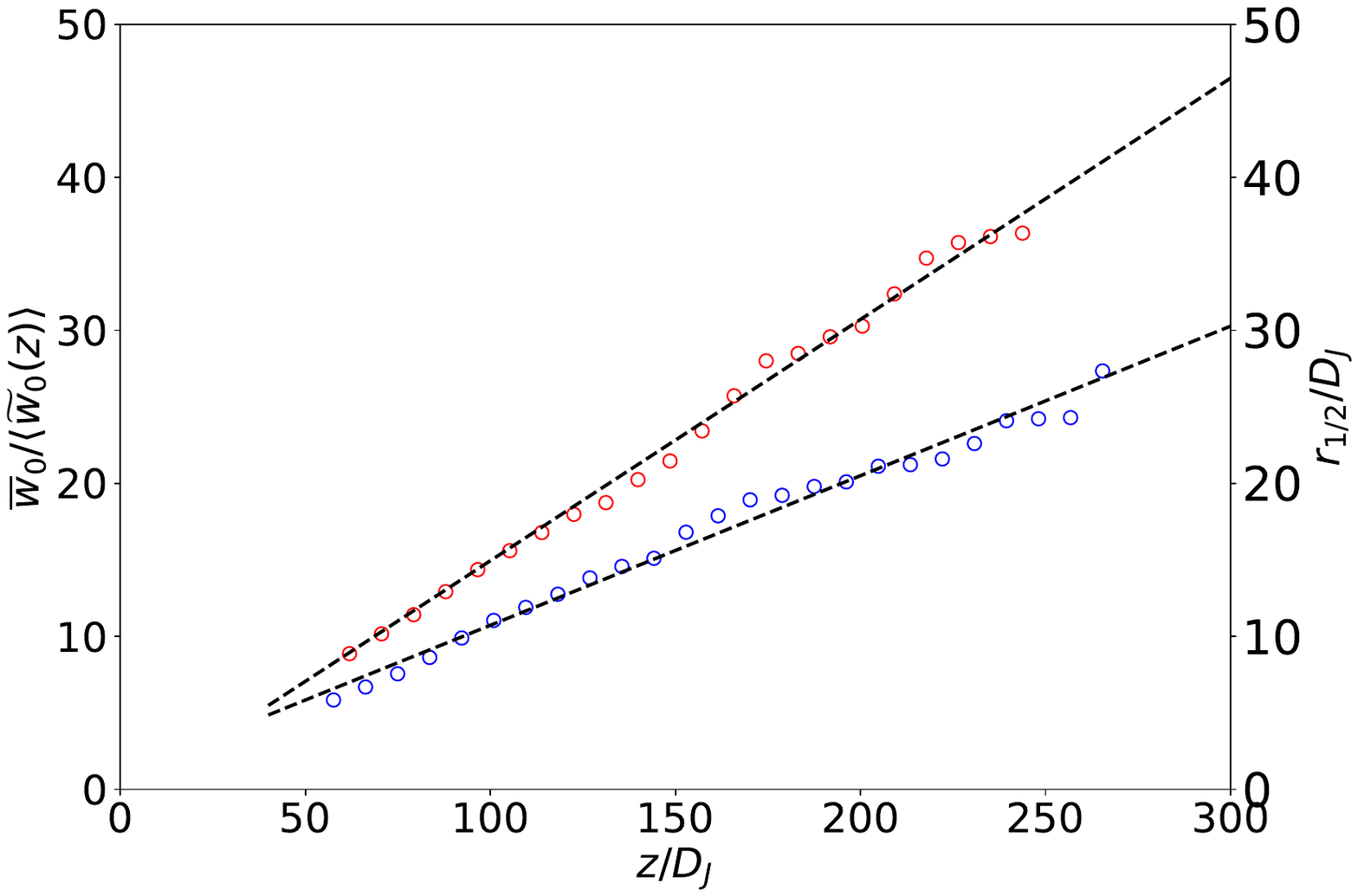}
\caption{Downstream variation of half width of the jet (\protect\bcircle, right axis) and the evolution of the inverse of the averaged centerline velocity (\protect\sq, left axis) from LES. The linear fit to the data is depicted by the black dashed line (\protect\blackdline). }
\label{fig:cent_evol}
\end{figure}

We first examine the centerline velocity $\langle\tilde{w}_0(z)\rangle$ and the jet half-width $r_{1/2}$ defined as usual according to 
\begin{equation}
    \langle \tilde{w}(z,r_{1/2}(z),0) \rangle = \frac{1}{2}\langle\tilde{w}_0(z)\rangle,
\end{equation}
where $z$ is the distance downstream of the experimental injection point. Figure \ref{fig:cent_evol} examines the evolution of the centerline velocity $\langle\tilde{w}_0(z)\rangle$ and half-width $r_{1/2}$ as a function of downstream distance scaled by the experimental nozzle diameter $D_J$. The injection velocity $\bw_0=11.8\;\mbox{ms}^{-1}$ at $z=0$ is used to scale the data. The inverse centerline velocity growth shown in figure \ref{fig:cent_evol} follows the expected hyperbolic law with a decay coefficient of $C_u = 6.3$  calculated from the slope of the curve. This is reasonably close to the value used in the 1D ODE model defined in equation (\ref{eqn:dissip_east})  and experimental data \cite{Panchapakesan1993,Hussein1994}.
The jet growth in the self similar region between $z = 50\;D_{J}$ and $z=300\;D_{J}$ is linear.  The slope of the curve, $S=0.097$ compares well with values obtained in the literature of $S\approx0.1$ \cite{Panchapakesan1993, Hussein1994} (see also equation \ref{eqn:dissip_east}). 
\begin{figure}
 \hspace{-1cm}
\centering
        \subfloat[Velocity\label{fig:ss_vel}]{\includegraphics[width=0.5\columnwidth]{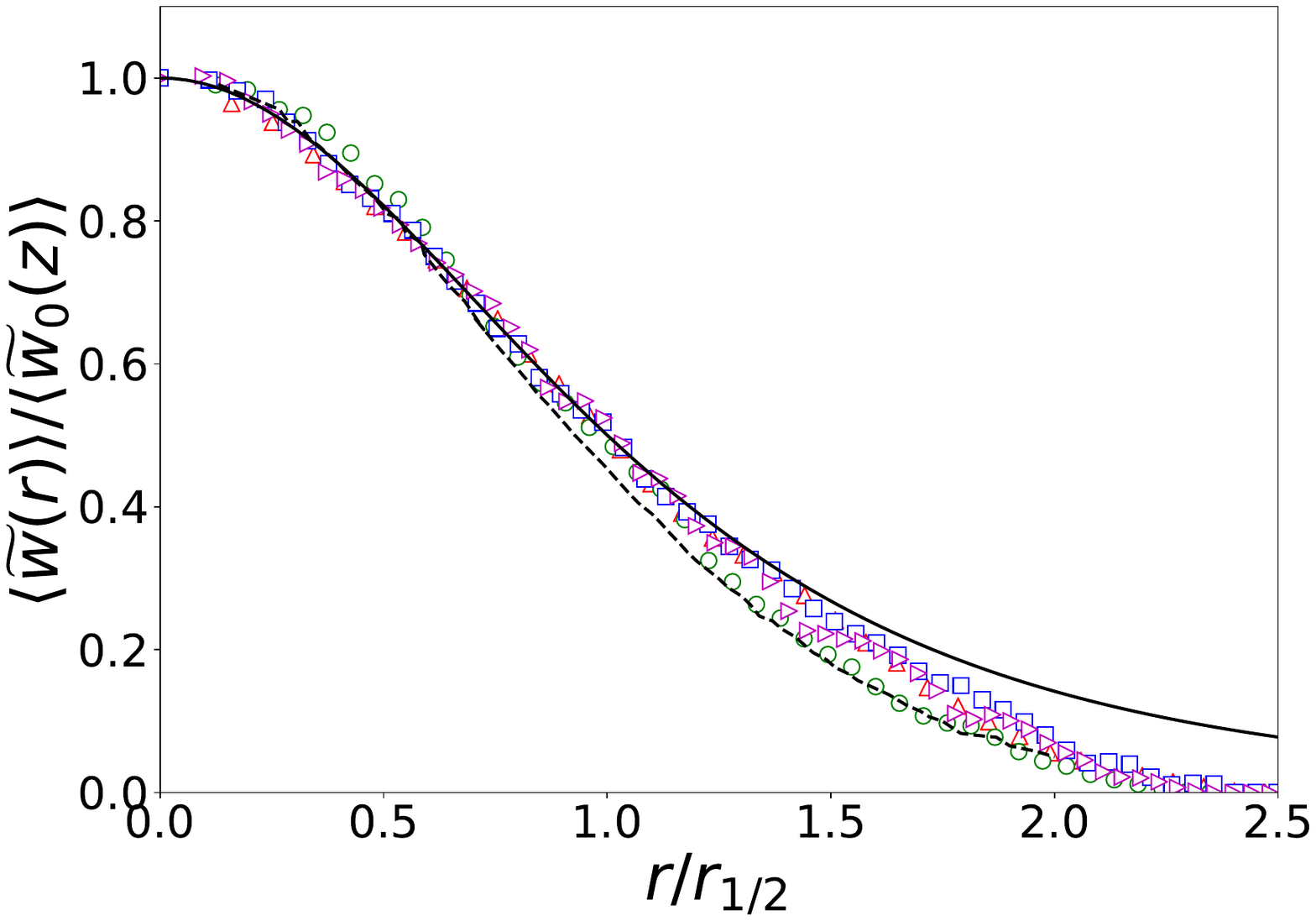}}
    \hfill
\subfloat[Concentration\label{fig:ss_conc}]{\includegraphics[width=0.5\linewidth]{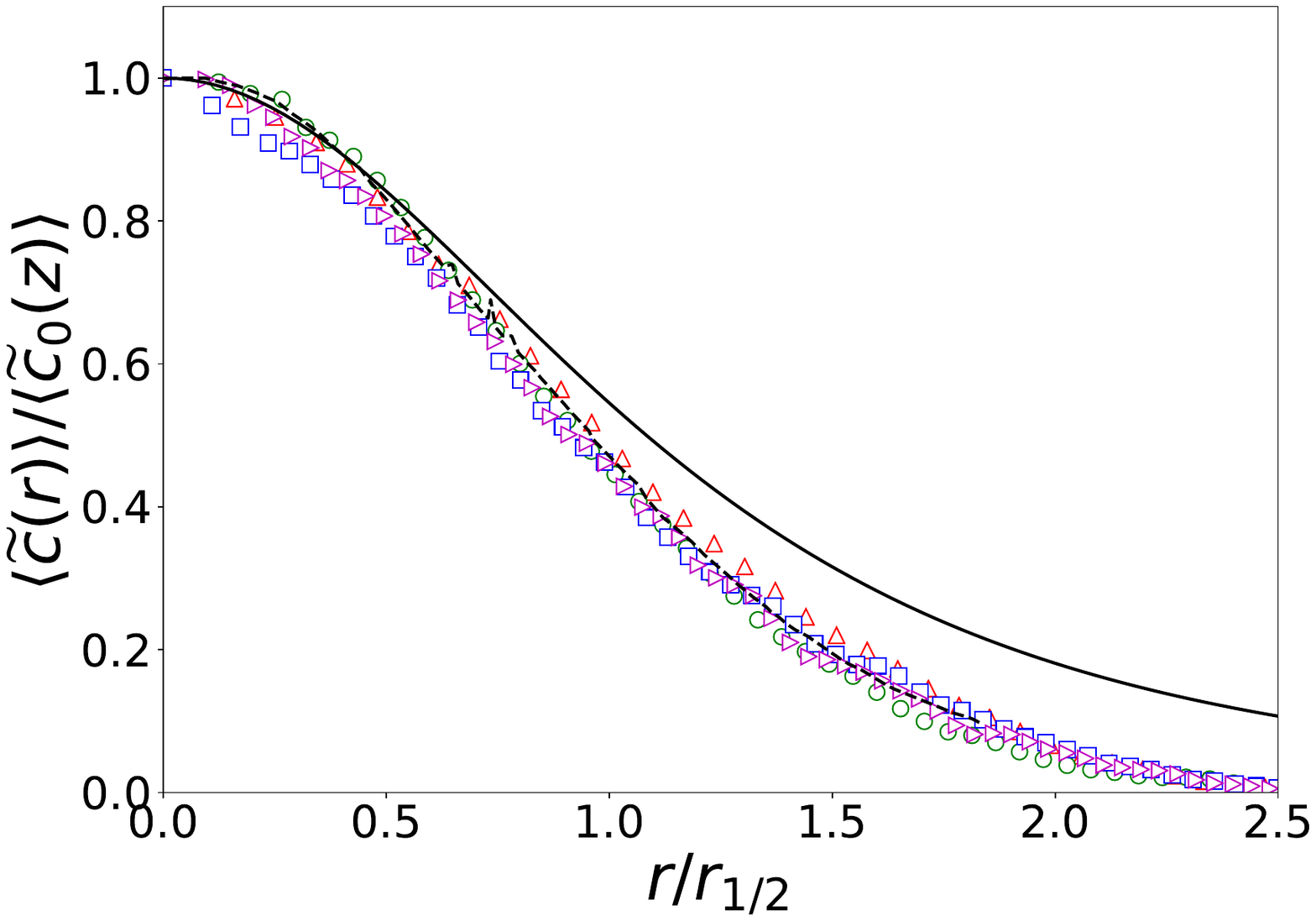}}
\caption{\protect\subref{fig:ss_vel} Averaged axial velocity profiles as function of normalized radial distance, \protect\subref{fig:ss_conc} Averaged concentration profiles at $z/D_{J} = 135$ (\protect\rtri), $z/D_{J} = 168$ (\protect\gcircle), $z/D_{J} = 211$ (\protect\bsq) and $z/D_{J} = 243$ ($\mtri$) as a function of self similarity variable $r/r_{1/2}$. The dashed line (\protect \blackldline) denotes the DNS data \cite{Lubbers2001} and the solid line (\protect\blackline) represents the analytical constant eddy-viscosity solution.}
\label{fig:sim_prof}
\end{figure}

We also document the radial distribution of velocity and concentration at different downstream locations, in figure \ref{fig:sim_prof}. The velocity profiles shown in figure \ref{fig:ss_vel} show approximate collapse on self similar behavior when normalized by the centerline value and plotted as a function of $r/r_{1/2}$, the radial coordinate scaled by the jet half-width. Additionally it shows good agreement with the constant eddy-viscosity similarity solution defined in equation (\ref{eqn:sim_vel}) in the central part of the jet, whereas it falls below the constant eddy viscosity solution at larger $r$ values, a behavior typically ascribed to the decreasing eddy viscosity in the outer parts of the jet \cite{Pope2001}. The DNS result from \citet{Lubbers2001}, shown as the black dashed lines, agrees well with LES data also in the outer portions of the jet.  The radial profiles of the total oil concentration normalized by the centerline value at various downstream locations is shown in \ref{fig:ss_conc}. Similar to the velocity profiles, the total concentration appears to be self similar when plotted as a function of $r/r_{1/2}$. Additionally, we plot the concentration profile derived from the constant eddy-diffusivity hypothesis, defined in equation (\ref{eqn:sim_conc}) with a Schmidt number, $Sc_T = 0.7$ as the solid black line.  We see that the analytic solution shows agreement with the simulation results near the centerline of the jet, with discrepancies at $r/r_{1/2} > 0.5$. Conversely, the data is in excellent agreement with the DNS data \cite{Lubbers2001} across the jet width. 

\begin{figure}
        \centering

        \includegraphics[width=0.6\linewidth]{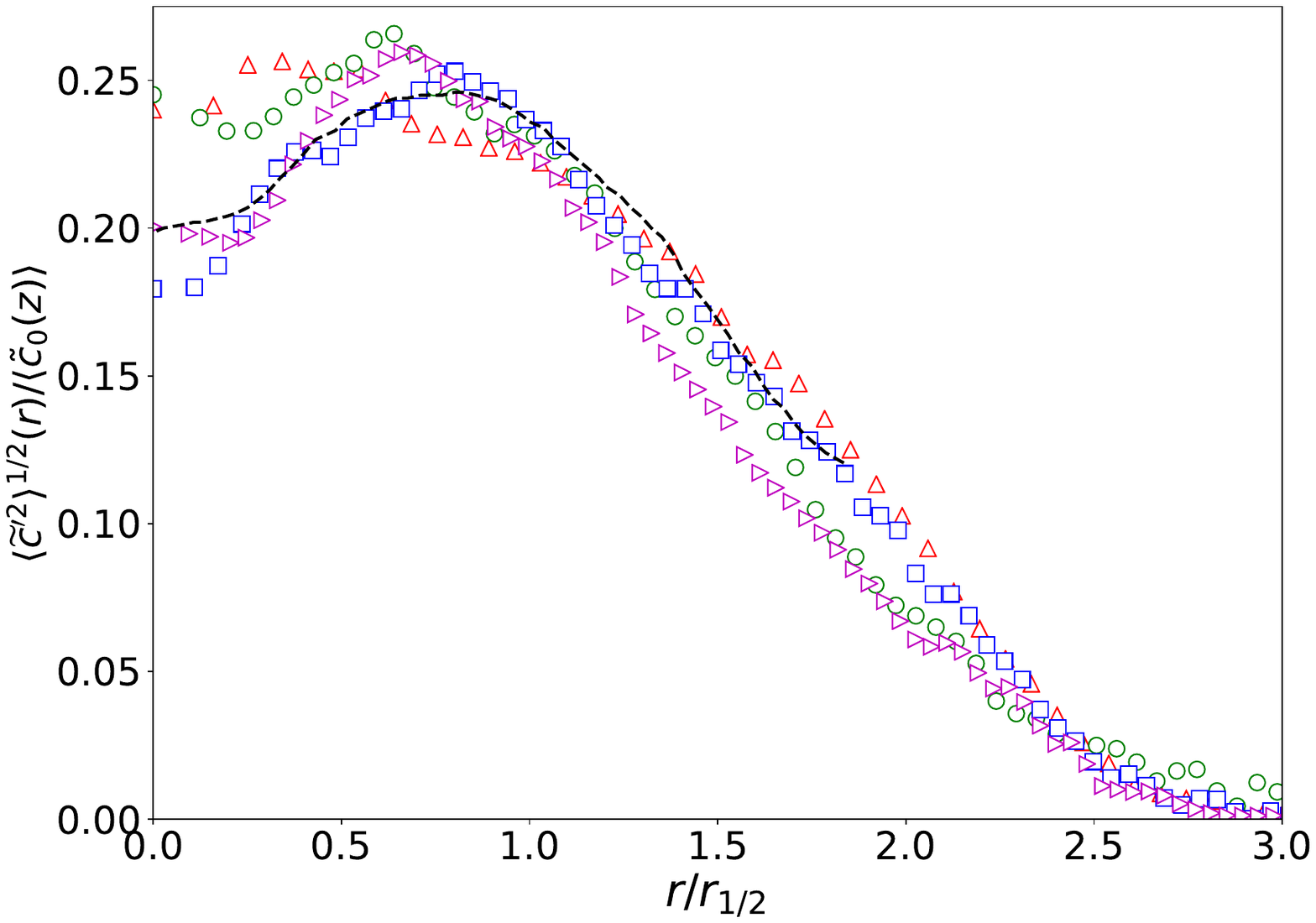}

  \caption{Radial distributions of concentration fluctuation root-mean-square at $z/D_{J} = 135$ (\protect\rtri), $z/D_{J} = 168$ (\protect\gcircle), $z/D_{J} = 211$ (\protect\bsq) and $z/D_{J} = 243$ ($\mtri$)), normalized by centerline mean concentration, as a function of   $r/r_{1/2}$.}
    \label{fig:RMS_conc}
\end{figure}

The radial distribution of the concentration fluctuations root-mean-square (r.m.s.) normalized by the mean centerline concentration is shown in figure \ref{fig:RMS_conc}. As  observed in prior simulations \cite{Lubbers2001}, the concentration fluctuation r.m.s.  shows an off-axis peak and in general shows good agreement with the DNS results of \citet{Lubbers2001}. 

\subsection{Droplet size distribution}

\begin{figure}
    \centering
    \subfloat[ $d = 3000\;\mu\mbox{m}$\label{fig:cont19}]{\includegraphics[width=0.25\linewidth,trim=13 13 16 13,clip]{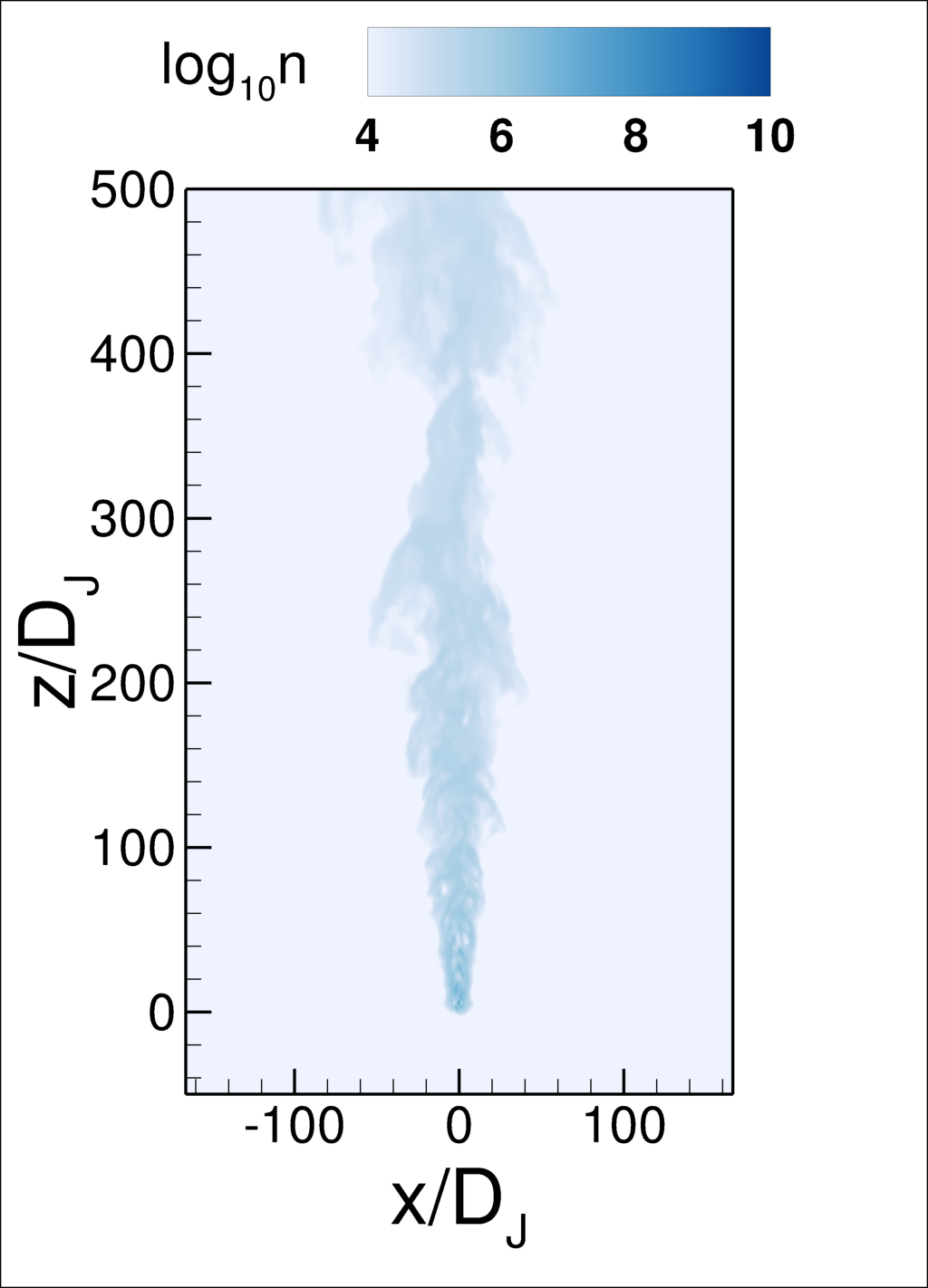}}
    \subfloat[ $d = 730\;\mu\mbox{m}$\label{fig:cont14}]{\includegraphics[width=0.25\linewidth,trim=8 8 16 8,clip]{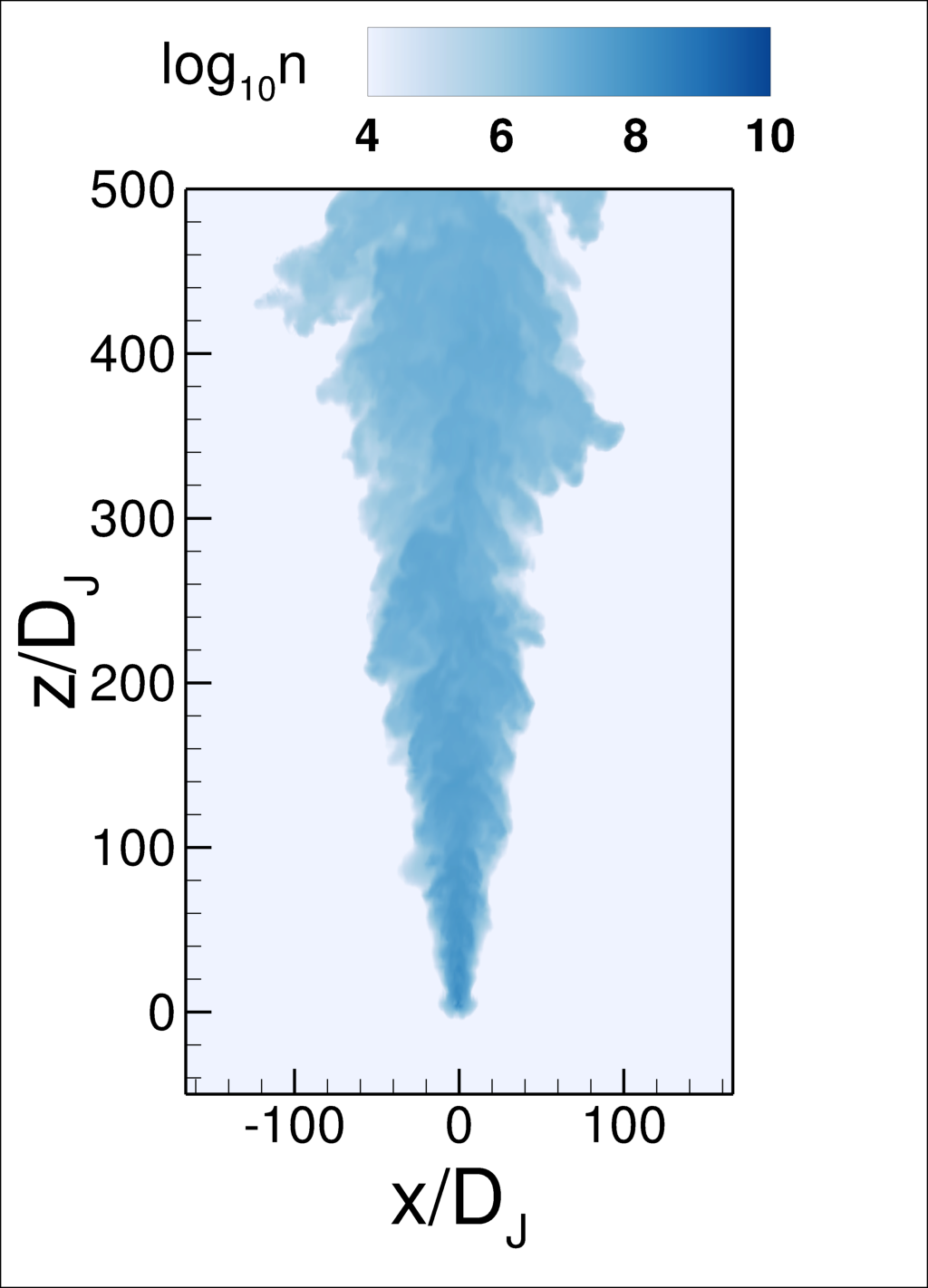}}
    \subfloat[ $d = 134\;\mu\mbox{m}$\label{fig:cont8}]{\includegraphics[width=0.25\linewidth,trim=8 8 16 8,clip]{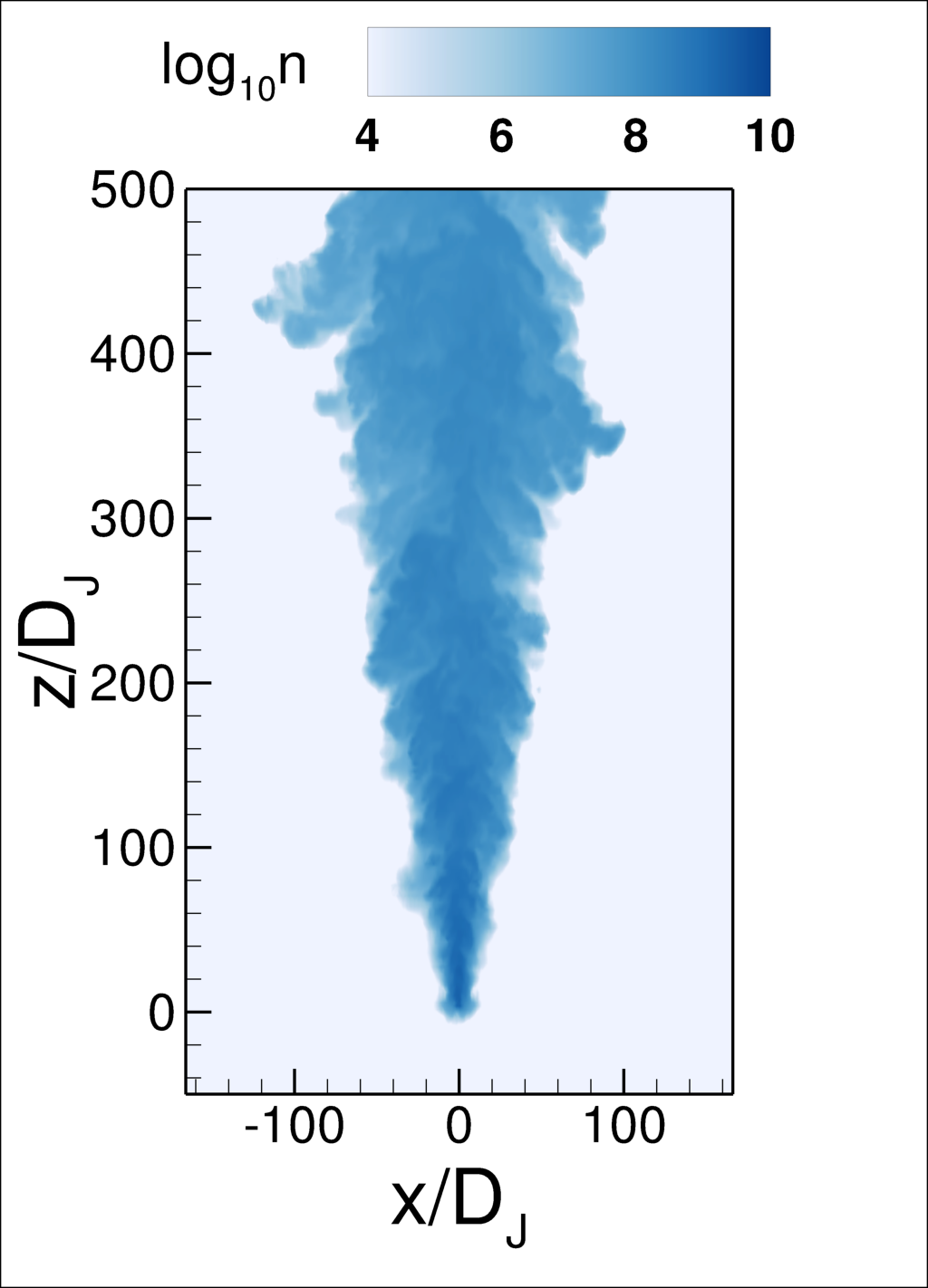}}
    \subfloat[ $d = 18\;\mu\mbox{m}$\label{fig:cont2}]{\includegraphics[width=0.25\linewidth,trim=8 8 16 8,clip]{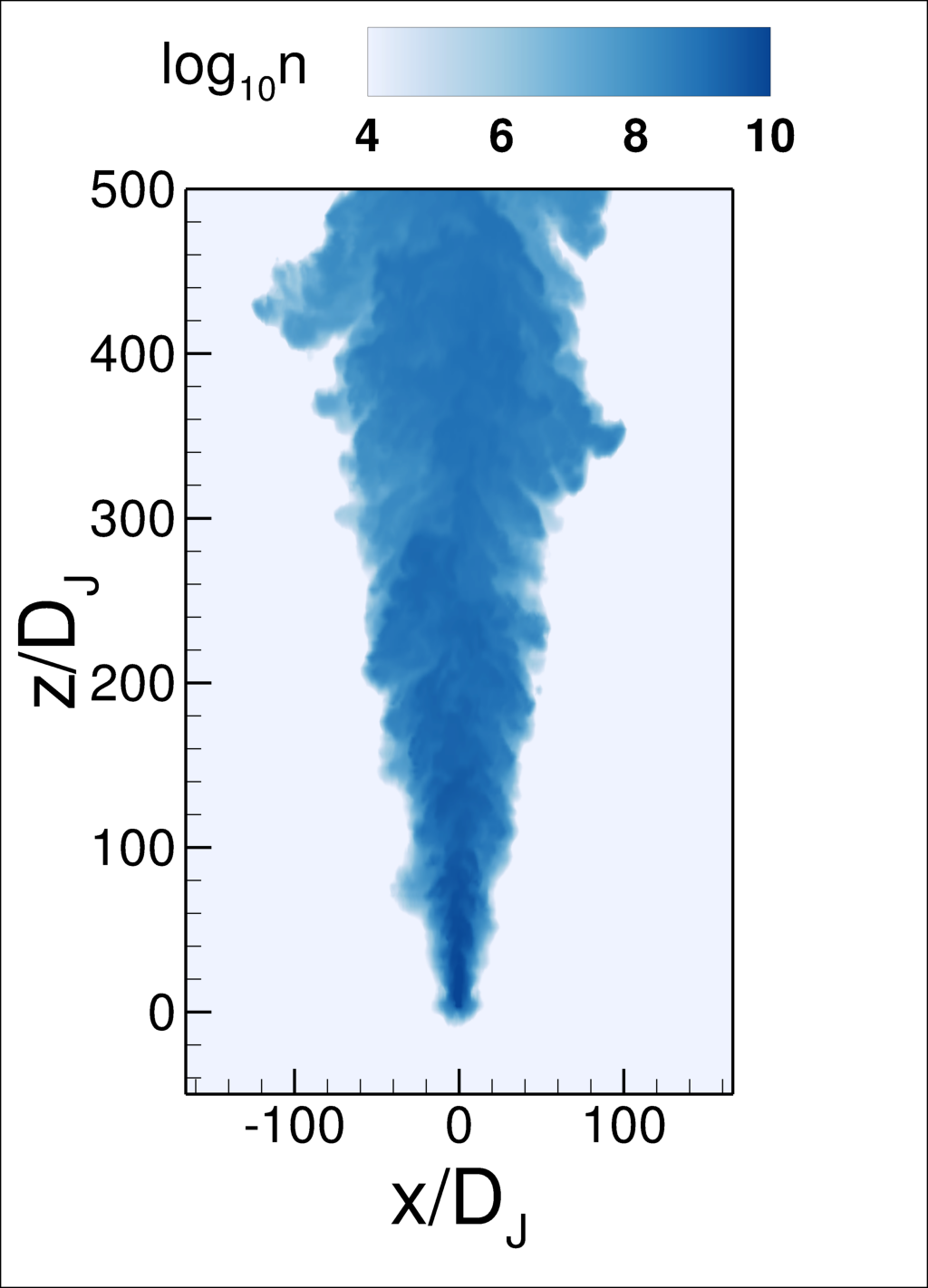}}
    
    \caption{Instantaneous snapshots of concentration fields at the midplane of the jet plotted in logarithmic scale for different droplet sizes. The domain has been cropped at $z/D_J = 500$ for visualization purposes. } 
    \label{fig:conc_c}
\end{figure}

Figure \ref{fig:conc_c} shows contour plots of instantaneous number density in logarithmic scale ($log_{10}(\tilde{n}_i)$) for four representative droplet sizes on the mid y-plane as a function of x and z. The concentration of the largest droplet size is in figure \ref{fig:cont19} and the smallest in figure \ref{fig:cont2}. We can see that far away from the nozzle the concentration of the largest size has decreased significantly due to breakup into the smaller droplet sizes. High concentrations for the smaller sizes can be seen to occur already in the near nozzle region due to the high dissipation rate that causes rapid droplet breakup there. 

\begin{figure}
        \centering
                \includegraphics[width=0.5\linewidth]{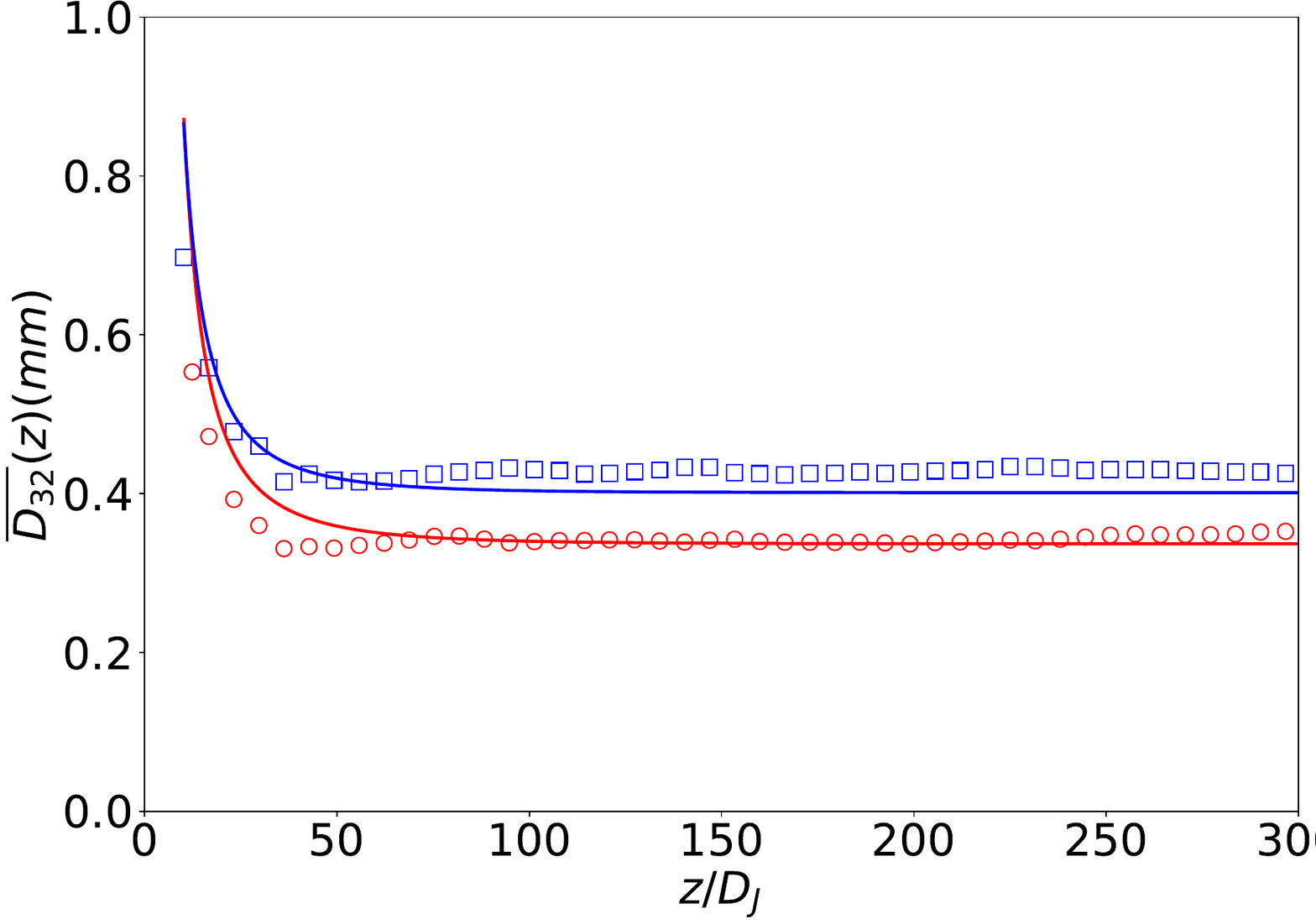}

 \caption{Evolution of Sauter mean diameter, $D_{32}$ as a function of downstream distance from the nozzle for SIM 1 (\protect \bsq) and SIM 2 (\protect \rcircle). The $D_{32}$ curves from the ODE model for both cases are depicted by a solid line. }
 \label{fig:d32_z}
\end{figure}
The Sauter mean diameter, ($D_{32})$ is often used  to quantify the size distribution by defining a characteristic diameter for a polydisperse distribution.  It is defined as the volume to surface area ratio of the distribution and is calculated from LES results using 
\begin{equation}\label{eqn:d32}
    \overline{D}_{32}  = \left<\frac{\sum_i \tilde{n}_id_i^3}{\sum_i \tilde{n}_id_i^2}\right>.
\end{equation}{}
 The Sauter mean diameter is calculated locally and at every time step using the instantaneous LES concentration and then averaged in time and polar direction $\theta$.
We plot the average $\overline{D}_{32}$ for the two simulations as a function of downstream distance in figure \ref{fig:d32_z}. 

The solid lines depict the results from the 1D ODE model. We see good agreement between the mean diameter calculated by the model and LES. Increasing Weber number reduces the overall mean diameter 

due to increased breakup frequency of the larger droplets in the near nozzle region. This reduction in mean diameter is reproduced to similar degrees in the LES results and the 1D ODE model.
We can see from \ref{fig:d32_z} that there is no significant change in the centerline mean diameter beyond $z = 100\;D_{J}$. This suggests that beyond this downstream position, no significant droplet breakup occurs. It is therefore sufficient to compare the results from the LES with that of the ODE model at $z/D_{J} = 333$ corresponding to a distance of $z=1\;m$ from the experimental nozzle. This allows us to save computational cost in the LES by limiting the analysis region only up to $z/D_{J} = 333$. We recall that the 1D ODE model has  been validated with the experimental data at $z/D_{J} = 666$ ($z=2\;m$) in figure \ref{fig:dsd_3_5} and showed very good agreement.

\begin{figure}
        \centering
    
        \subfloat[SIM 1\label{fig:sim1_les_ode}]{\includegraphics[width=0.5\linewidth]{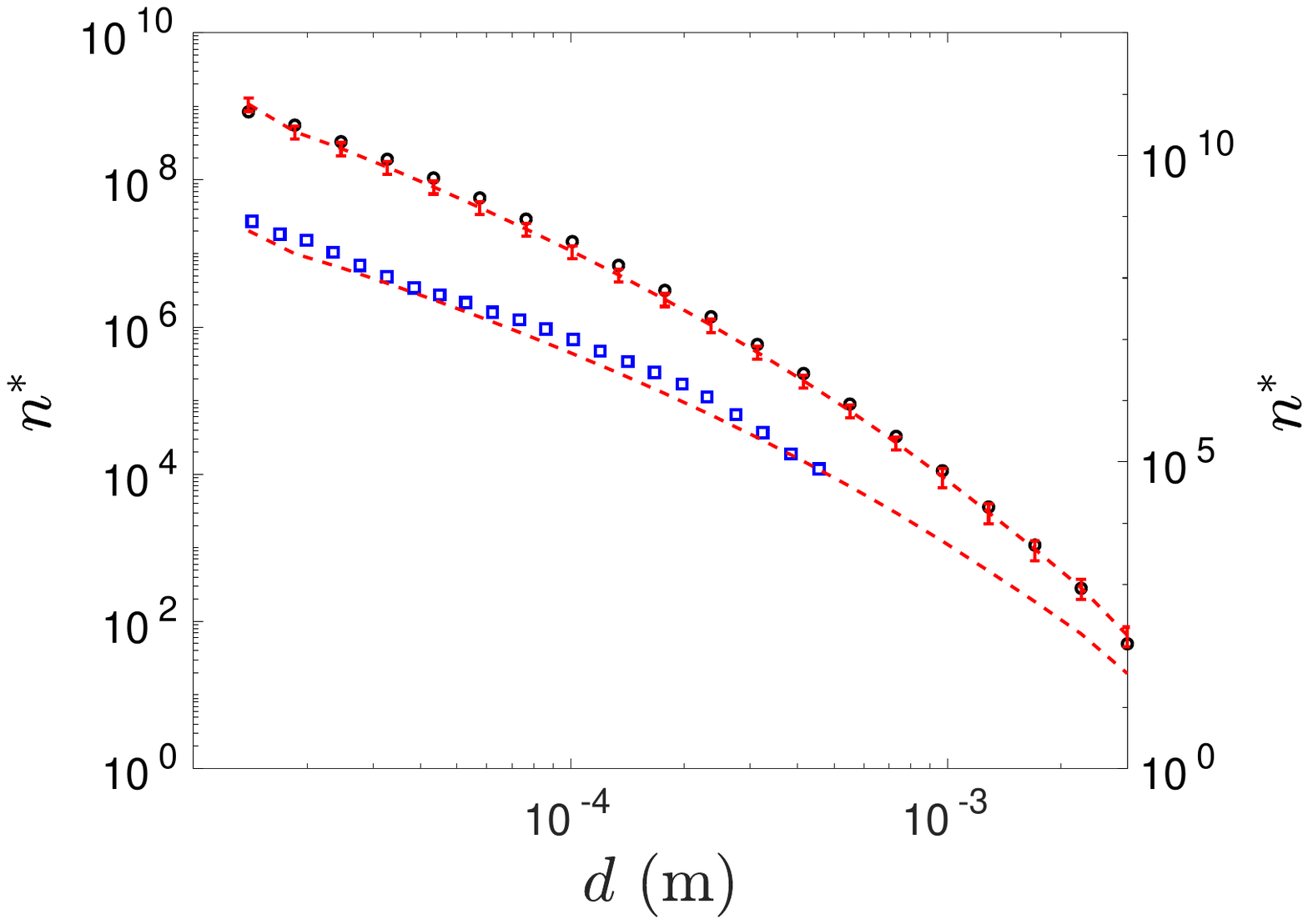}}
\hfill
 \subfloat[SIM 2\label{fig:sim2_les_ode}]{\includegraphics[width=0.5\linewidth]{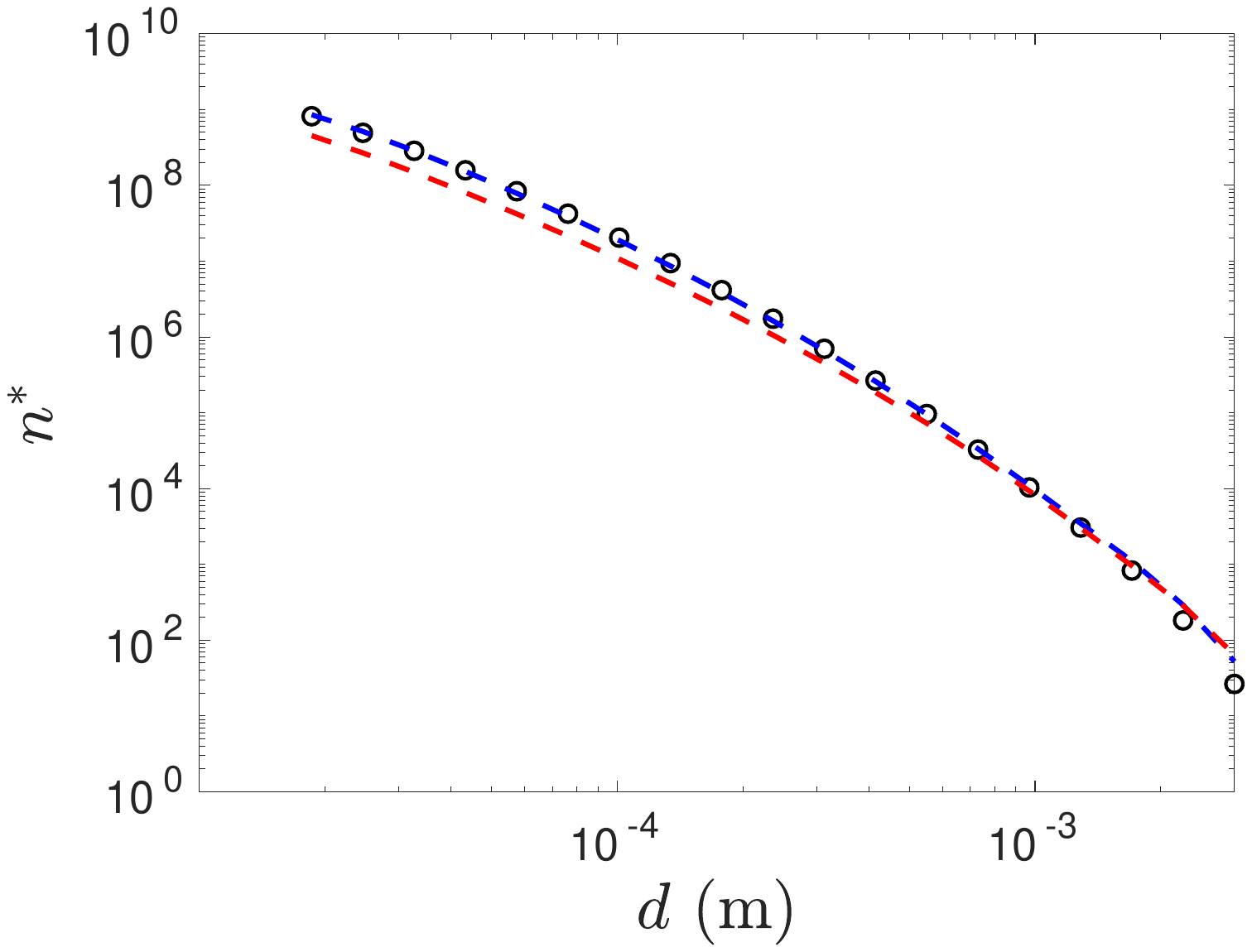}}
  \caption{(a) Comparison of centerline droplet size distribution 
  from experimental data (\protect\bsq, right axis)
  at $z/D_J=666$ 
  with extended LES results (also right axis, \protect\reddline). The latter is obtained by solving Eq. (\ref{eqn:centerline_conc}) using the LES data as inlet condition at $z/D_J=333$ (left axis) as initial condition (these LES data at $z/D_J=333$ are shown by the top \protect\reddline  line). Error bars display the r.m.s. at $z/D_J=333$ due to turbulence.  The 1D ODE model applied between $z/D_J=2$ to 333 is depicted by (\protect\kcircle, left axis).   (b) Comparison of droplet size distribution from SIM 2 (\protect\blueddline) with 1D ODE model (\protect\kcircle) and SIM 1 size distribution (\protect\reddline) at $z/D_J = 333$.}
    \label{fig:LES_ODE_DSD}
\end{figure}

 The LES number concentration fields are averaged in time and the droplet size distribution $\langle{n^*}\rangle$, is calculated using equation ($\ref{eqn:num_dens}$) by normalizing the number concentration by the bin width.  Figure \ref{fig:LES_ODE_DSD} compares the size distribution at the centerline at $z/D_{J}=333$ to the size distribution obtained from the 1D ODE model at the same location. We can see that the LES and the 1D ODE model provide very consistent predictions of the size distribution at the centerline.  The error bars provide additional information regarding the turbulent fluctuations of $\langle{n^*}\rangle$: they are calculated using the root-mean-square (r.m.s) of the concentration for each droplet size at the centerline. 
  Beyond  $z/D_{J} = 333$ the evolution of each of the bins concentration is affected only by transport and not by breakup. 
  This allows us to carry out an additional validation step by using equation (\ref{eqn:centerline_conc}) with $S_{b,i}=0$ to calculate the thus extrapolated LES size distribution  at $z/D_{J} = 666$. 
The resulting distribution from the extrapolated LES is compared with the experimental measurements in figure \ref{fig:LES_ODE_DSD} (blue squares compared to red dashed line), with  excellent agreement. 
 
 The size distribution for the case with increased Weber number is shown in figure \ref{fig:sim2_les_ode}. We can see that due to increased breakup of the larger sized droplets, the number density of the smaller-diameter bins is larger, and the distribution has  a higher slope throughout. This effect is also observed in experiments when dispersant is premixed with oil \cite{Brandvik2013,Zhao2016evolution,Murphy2016,li2017}. This shift of the concentration towards the smaller scales results in the lower Sauter mean diameter observed in figure \ref{fig:d32_z} for SIM 2.

 \begin{figure}
        \centering
\subfloat[SIM 1\label{fig:cent_conc_sr2}]{\includegraphics[width=0.5\linewidth]{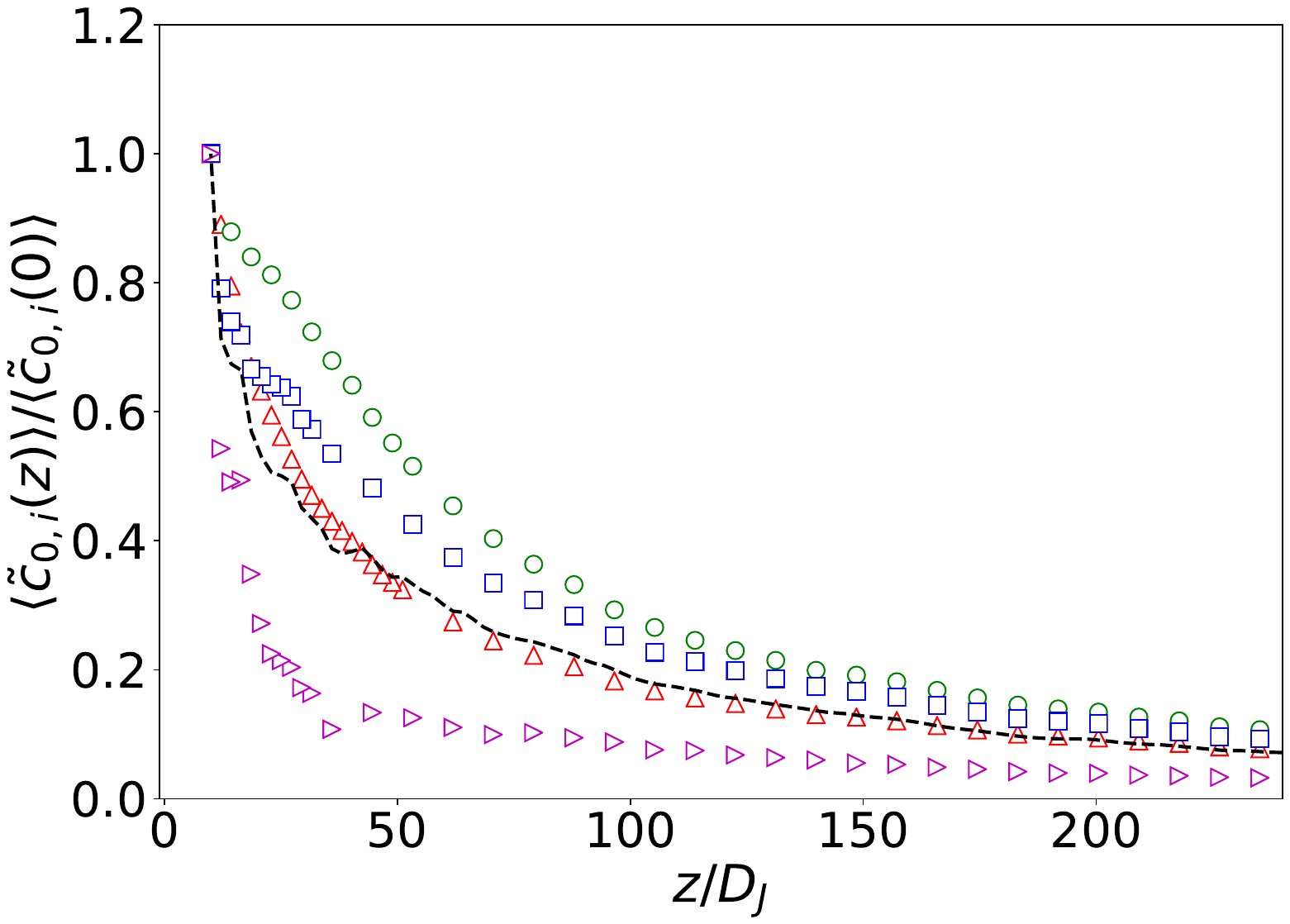}}
    \hfill
    \subfloat[SIM 2\label{fig:cent_conc_we}]{\includegraphics[width=0.5\linewidth]{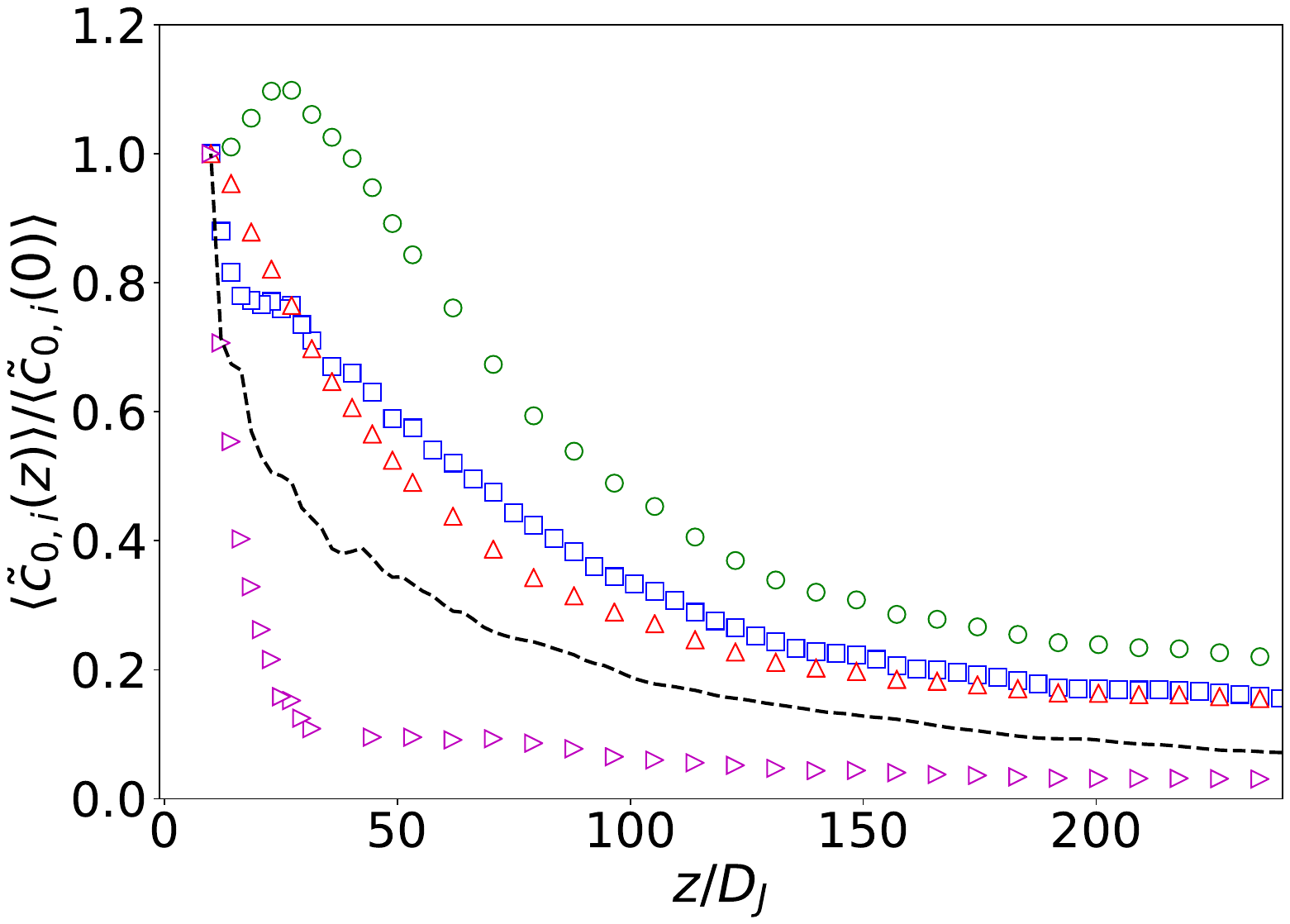}}
    
  \caption{Decay of centerline concentration of different droplet sizes as a function of downstream distance from the nozzle. The symbols represent the LES evolution for (a) SIM 1 and (b) SIM 2. The symbols are $d = 14\; \mu \mbox{m} $ (\protect\rtri), $d = 100\; \mu \mbox{m} $ (\protect\gcircle), $d = 550\; \mu \mbox{m} $ (\protect\bsq) and $d = 3\;\mbox{mm} $ ($\mtri$). The  total concentration from SIM 1 and SIM 2 is represented by the black dashed line.}
    \label{fig:cent_conc}
\end{figure}

\begin{figure}
\centering
\includegraphics[width=0.6\linewidth]{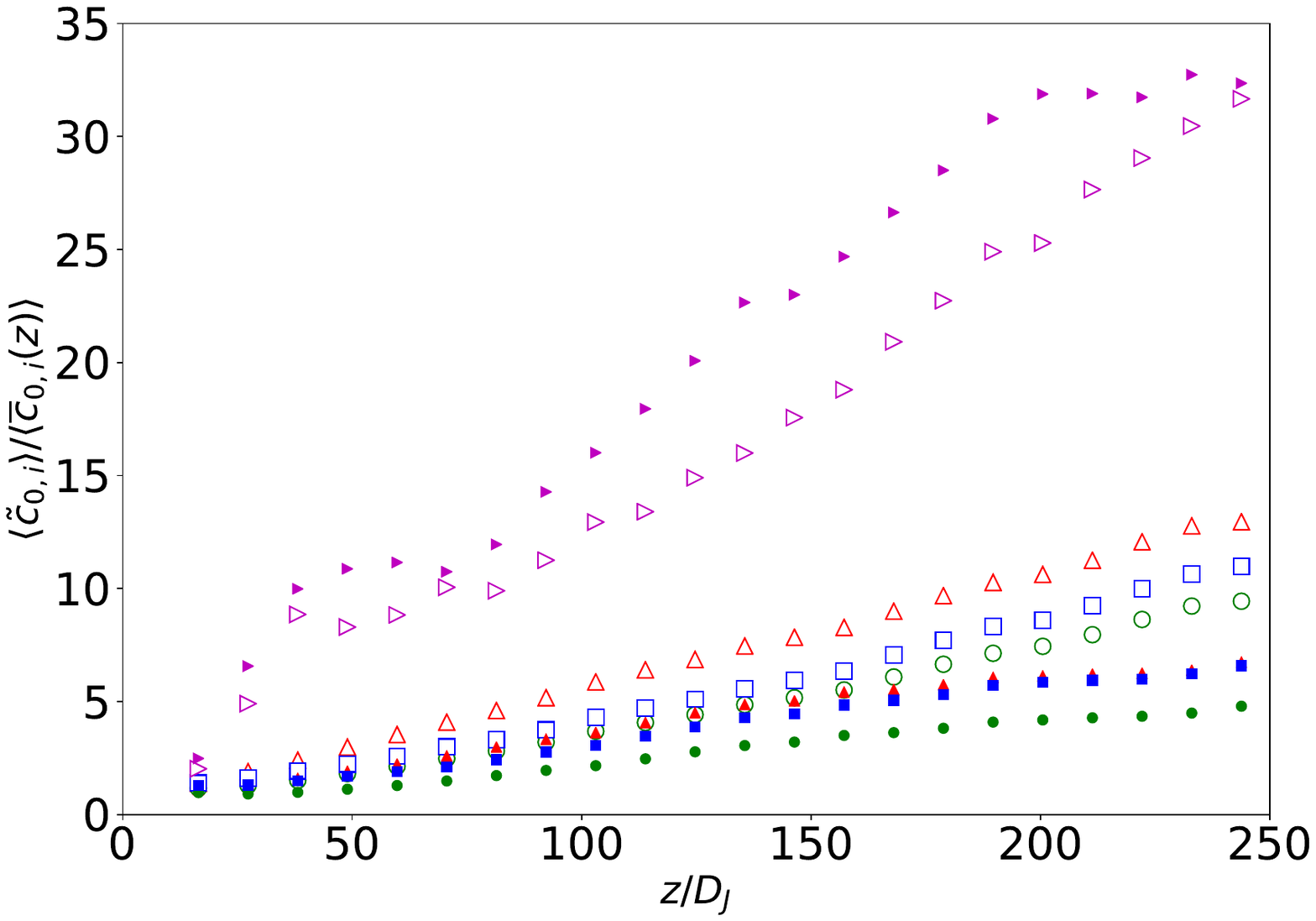}

\caption{Evolution of the inverse centerline concentration for SIM 1(open symbols) and SIM 2 (closed symbols). The symbols are $d = 14\; \mu \mbox{m} $ (\protect\rtri), $d = 100\; \mu \mbox{m} $ (\protect\gcircle), $d = 550\; \mu \mbox{m} $ (\protect\bsq) and $d = 3\;\mbox{mm} $ ($\mtri$).  }
\label{fig:inv_conc}
\end{figure}
 
 LES allows us to analyze the evolution of the droplet plumes for each droplet size. The effects of breakup are clearly visible in figure \ref{fig:cent_conc}. For the largest droplet size, $d=3\;\mbox{mm}$ we can see a rapid decay in the breakup dominated zone, approximately $z < 50\;D_{J}$ after which the change in concentration is primarily transport dominated. The change of initial slope and shape of the profiles among different droplet sizes is non-monotonic. The smallest droplets, of  size $d = 14\;\mu\mbox{m}$ do not break down further and its bin acts as a sink for all the other sizes, resulting in a  concentration profile that appears smoother and more monotonic than the other bins' mean concentration. The effect of increasing Weber number is to increase the rate of breakup of the larger droplets   due to the reduction in surface tension, leading to the increase in concentration of the smaller droplet sizes as can be seen from figure \ref{fig:cent_conc_we}.  Intermediate sized droplets behave as both a source, breaking up into smaller droplets and a sink, where larger droplets break up into the intermediate ones. This trend can be observed from the profile of the $d=550\;\mu\mbox{m}$ droplet in figure \ref{fig:cent_conc_we} that shows a peak near the nozzle followed by a decay of concentration. The profiles of the total concentration (summed over all bins), $\tilde{c}_0(z)$ for both the simulations are shown in figure \ref{fig:cent_conc} as dashed lines. As expected, we confirm that the evolution of the total concentration is fairly insensitive to Weber number. 

Figure \ref{fig:inv_conc} depicts the downstream evolution of the inverse of the centerline concentration for SIM 1 (open symbols) and SIM 2 (filled symbols) for different droplet sizes. The slope of the growth of the inverse concentration is size dependent, with a maximum slope for the largest droplet size, due to their rapid breakup. We can see that the change in slope for the different droplet sizes is non-monotonic, with the concentration of the  $d=550\;\mu m$ droplet decaying faster than the $d=100\;\mu m$ droplet concentration. Increasing Weber number results in a shallower slope for the smaller sizes as can be seen from the solid symbols. Conversely for the larger sizes (primarily acting as sinks), the growth of the inverse concentration is more rapid due to the increased breakup frequency.

 \subsection{Temporal variability of size distribution}
 
As discussed earlier, LES allows us to calculate the variability of the droplet size distribution that averaged integral models or RANS cannot obtain. We proceed to quantify the radial distributions of the mean and standard deviations of practically relevant quantities such as the Sauter mean diameter, the total surface area and the inverse droplet breakup time-scale.

\begin{figure}
        \centering
        \subfloat[SIM 1\label{fig:d32_s1}]{\includegraphics[width=0.5\columnwidth]{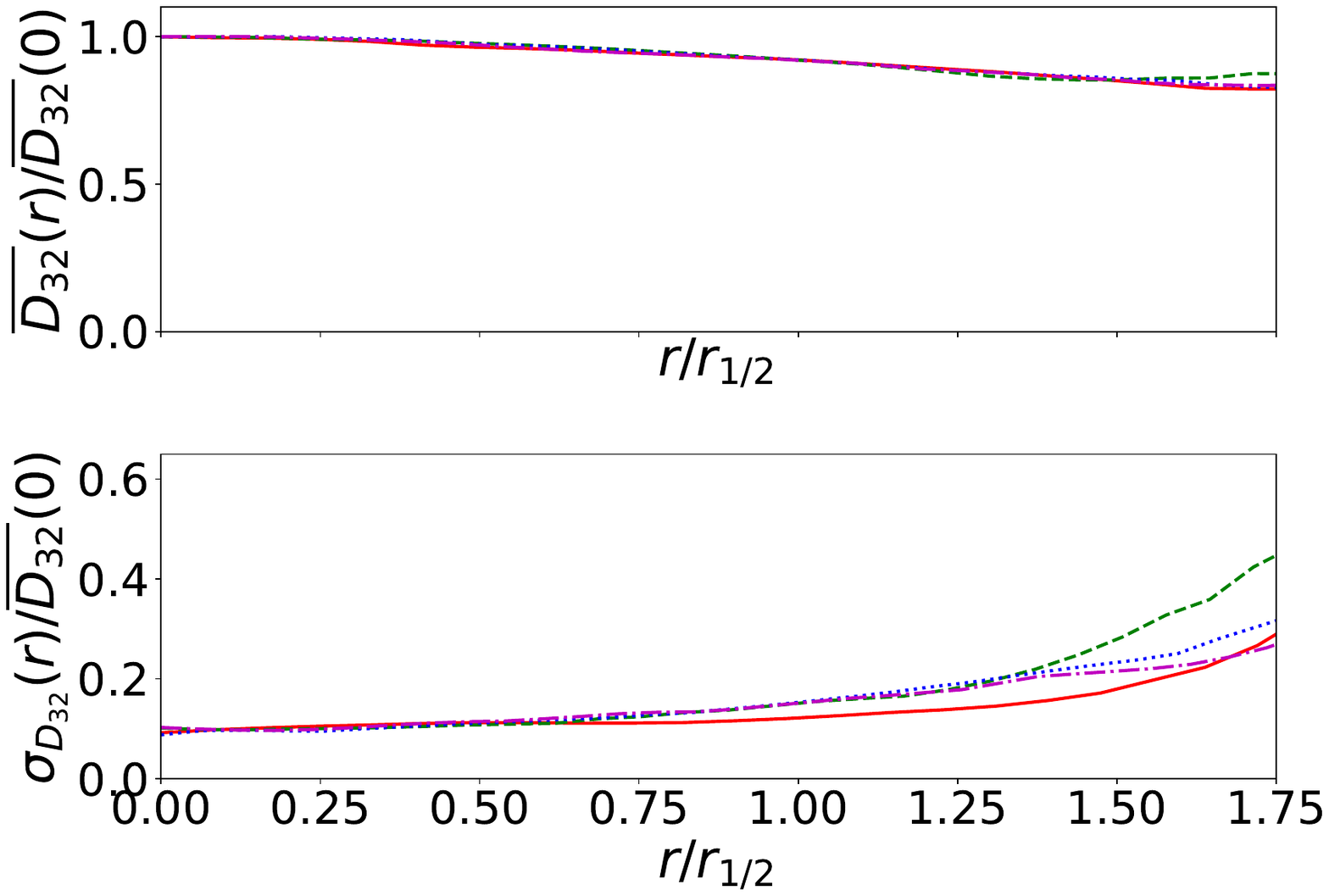}}
        \hfill
        \subfloat[SIM 2 \label{fig:d32_s2}]{\includegraphics[width=0.5\columnwidth]{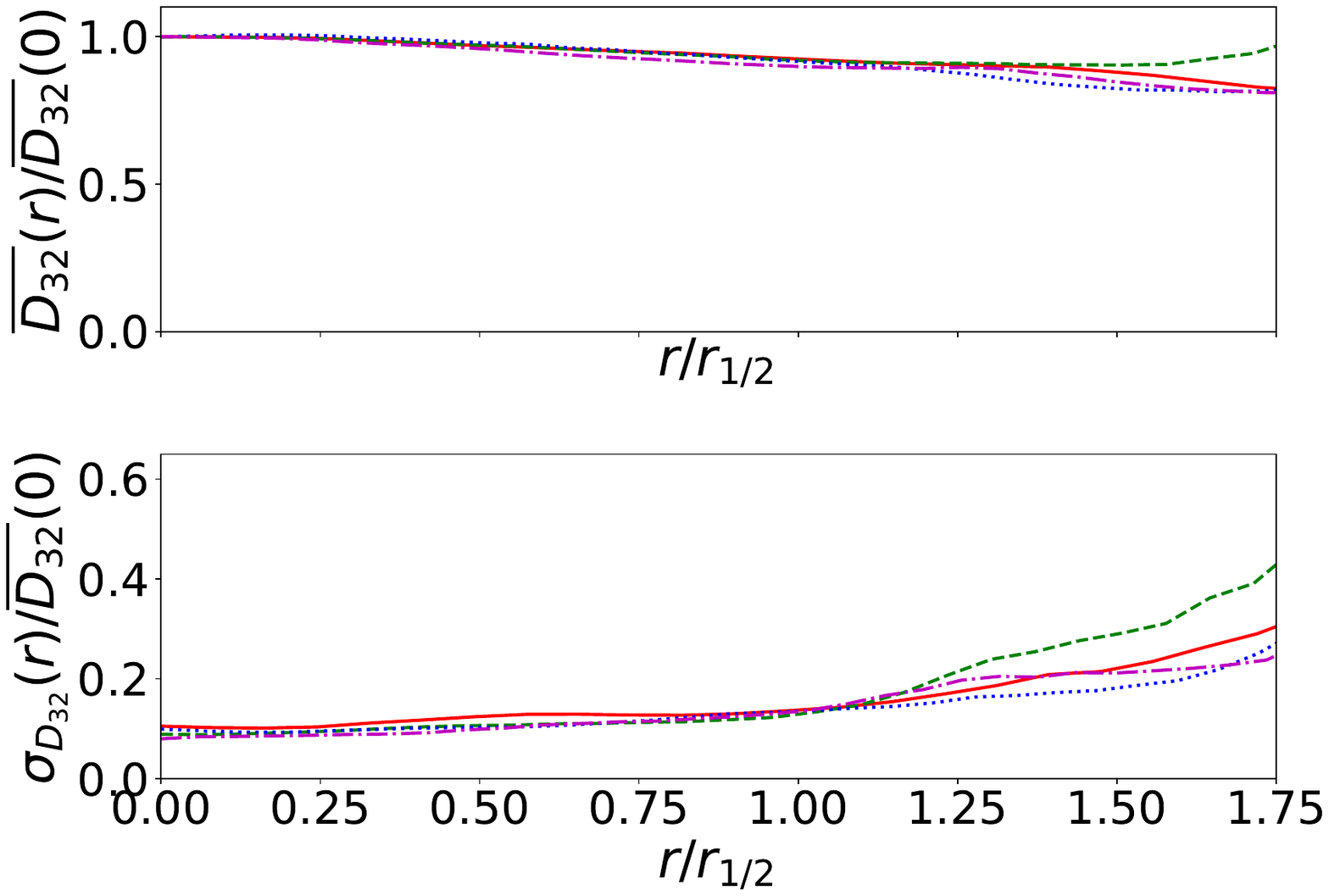}}
        
  \caption{The top panel depicts the radial distribution of the averaged Sauter mean diameter, $D_{32}$ normalized by its centerline value while the bottom panel depicts the normalized standard deviation at $z/D_{J}= 135$ (\protect\redline), $z/D_{J}= 168 $ (\protect\greendline), $z/D_{J}= 211$ (\protect\blueddline), $z/D_{J}= 243$ (\protect\magdline) for (a) SIM 1 and (b) SIM 2.}
    \label{fig:d32_pdf}
\end{figure}
We begin by examining the radial distribution of the mean diameter defined in equation (\ref{eqn:d32}) normalized by its centerline value at different downstream locations in figure \ref{fig:d32_pdf}. The mean diameter exhibits a weak decay with radial distance, with the centerline value decreasing by $20\%$ at $r/r_{1/2} =1.75$.
The standard deviation of the Sauter mean diameter, normalized by the mean diameter at the centerline, is relatively low near the center (around 10\%) but increases with radial distance towards the edge of the jet. We observe a maximum variability for the location farthest downstream from the nozzle. The increased variability at the edge of the jet can be attributed to the entrainment of fluid at the edges that results in increased dilution of concentration.
Increased variability in $D_{32}$ means that the size distribution displays changes as function of time at a particular location.  Increasing Weber number has minimal effect on the radial profiles of the normalized mean diameter. The variability is unchanged near the centerline of the jet but is slightly reduced towards the edge downstream of the nozzle.

\begin{figure}
        \centering
    
        \subfloat[SIM 1]{\includegraphics[width=0.5\linewidth]{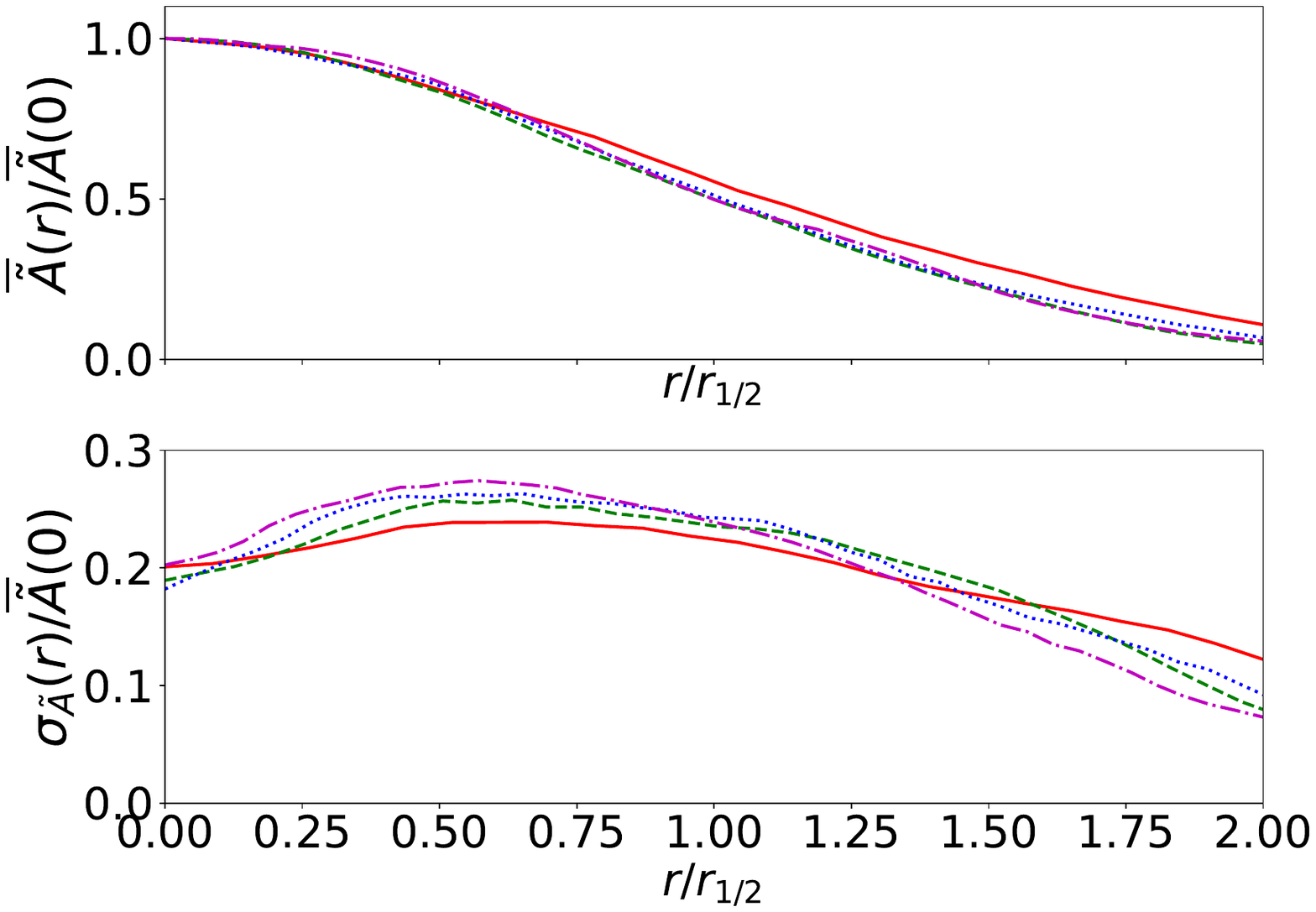}}
        \hfill
        \subfloat[SIM 2]{\includegraphics[width=0.5\linewidth]{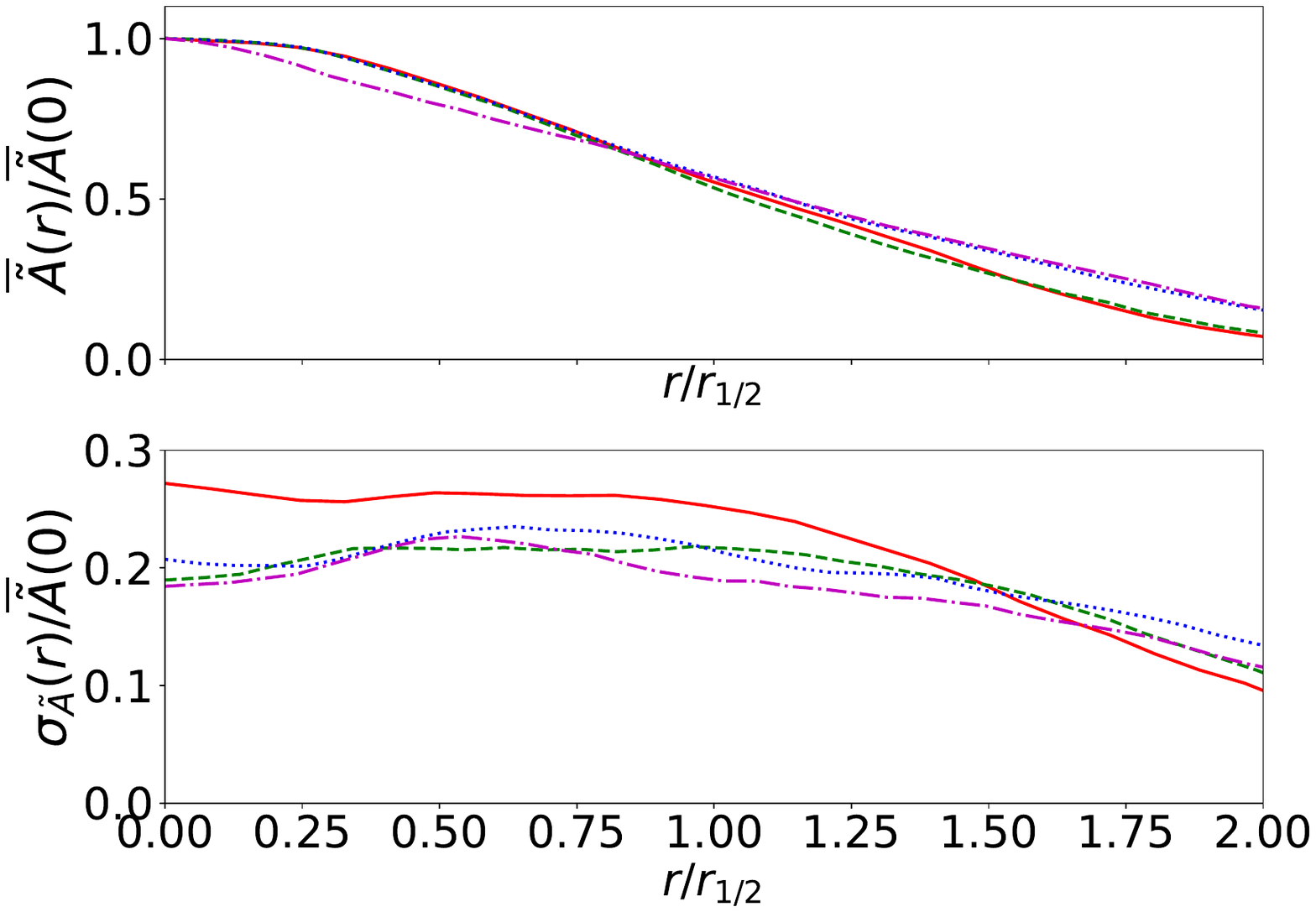}}
  \caption{The top panel depicts the radial distribution of the averaged total surface area, $\widetilde{A}$ normalized by its centerline value while the bottom panel depicts the normalized standard deviation at $z/D_{J}= 135$ (\protect\redline), $z/D_{J}= 168 $ (\protect\greendline), $z/D_{J}= 211$ (\protect\blueddline), $z/D_{J}= 243$ (\protect\magdline) for (a) SIM 1 and (b) SIM 2.. }
    \label{fig:tot_ar_pdf}
\end{figure}

Next, we examine the total surface area of the oil-water interface per unit volume of fluid, defined as:
\begin{equation}
\widetilde{A}({\bf x},t) = \sum_i \tilde{n}_i({\bf x},t) \, \pi d_i^2     
\end{equation}{}
This quantity is critical in determining reaction rates for processes that occur at the surface of the droplet. The radial profiles of the total area closely follow those of the mean concentration, and exhibit a reasonable collapse when plotted against the self similar co-ordinate. Interestingly, the temporal variability as quantified by means of the r.m.s. of $\widetilde{A}$ exhibits the opposite trend as compared to the Sauter mean diameter variability. There is maximum variation about the mean total area at about $r/r_{1/2} \sim 0.6$, which subsequently decays towards the edges. The shape of the profiles is similar to the concentration variance shown in figure \ref{fig:RMS_conc}. We see that the normalized profiles for the mean and standard deviation are relatively unchanged with changing Weber number.  Such information expands on that provided by  reduced or RANS type models, that are capable of quantifying only the mean of these quantities.  

\begin{figure}
        \centering
    
       \subfloat[SIM 1]{ \includegraphics[width=0.49\linewidth]{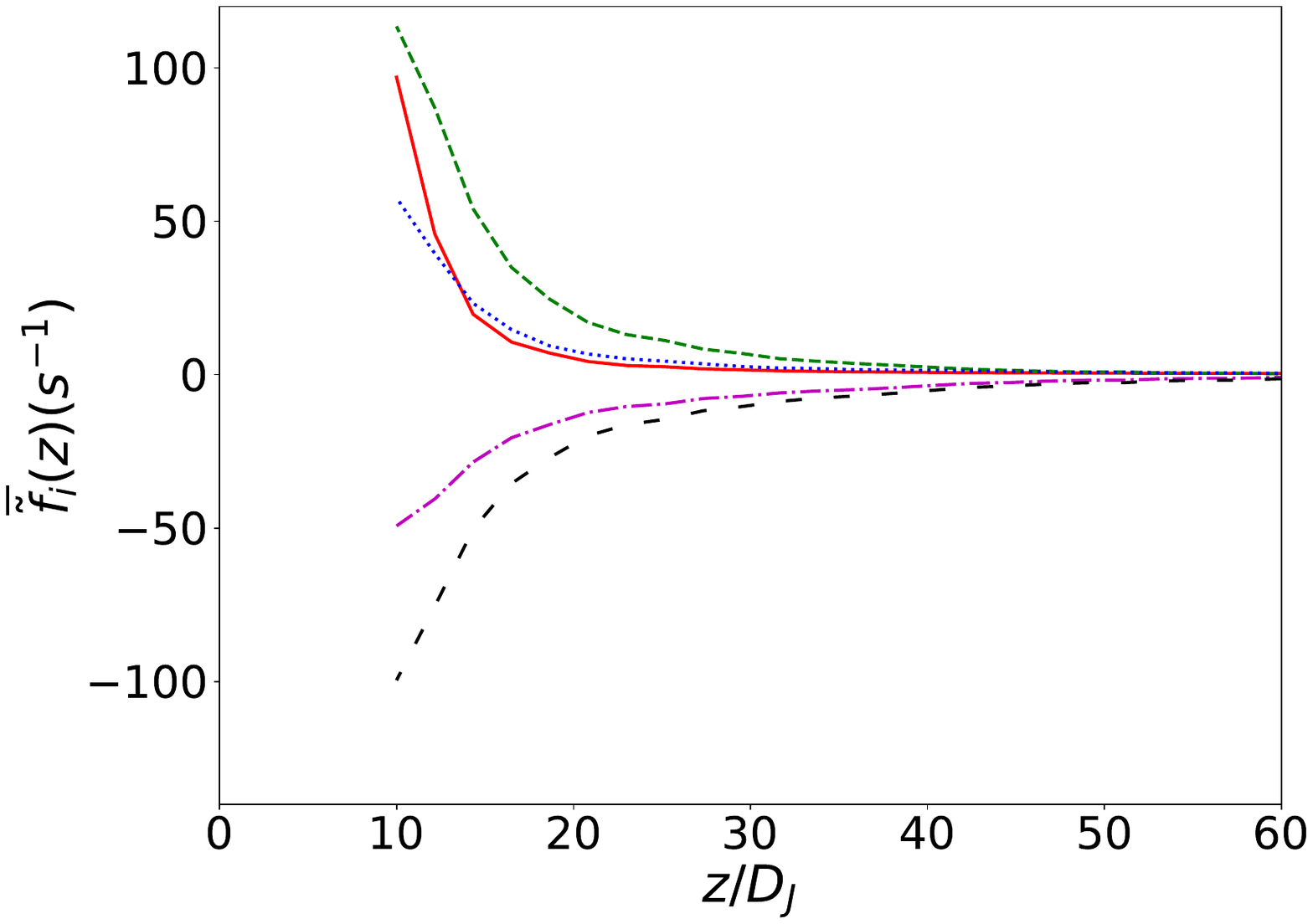}}
\hfill
    \subfloat[SIM 2]{ \includegraphics[width=0.49\linewidth]{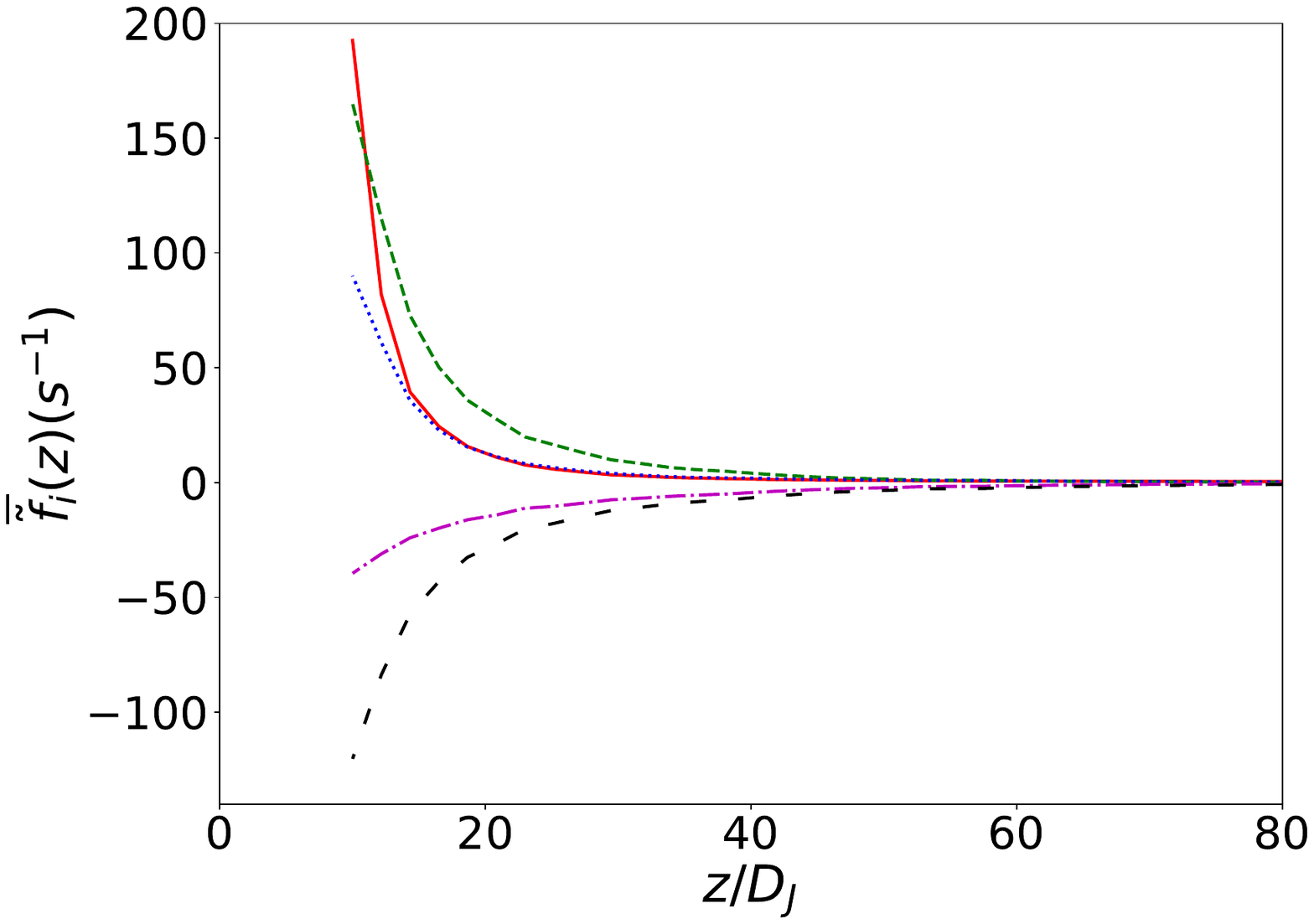}}
  \caption{ Evolution of the inverse breakup time scale with downstream distance for (a) SIM 1 and (b) SIM 2. The lines are $d = 14\;\mu \mbox{m}$ (\protect\redline), $d = 100\;\mu \mbox{m}$ (\protect\greendline), $d = 550\;\mu \mbox{m}$ (\protect\blueddline) and $d = 2261\;\mu \mbox{m}$ (\protect\magdline) and $d = 3000\;\mu \mbox{m}$ (\protect\blackldline).  }
    \label{fig:ts_z}
\end{figure}

\begin{figure}
        \centering
    
\subfloat[SIM 1\label{fig:ts_src}]{\includegraphics[width=0.5\linewidth]{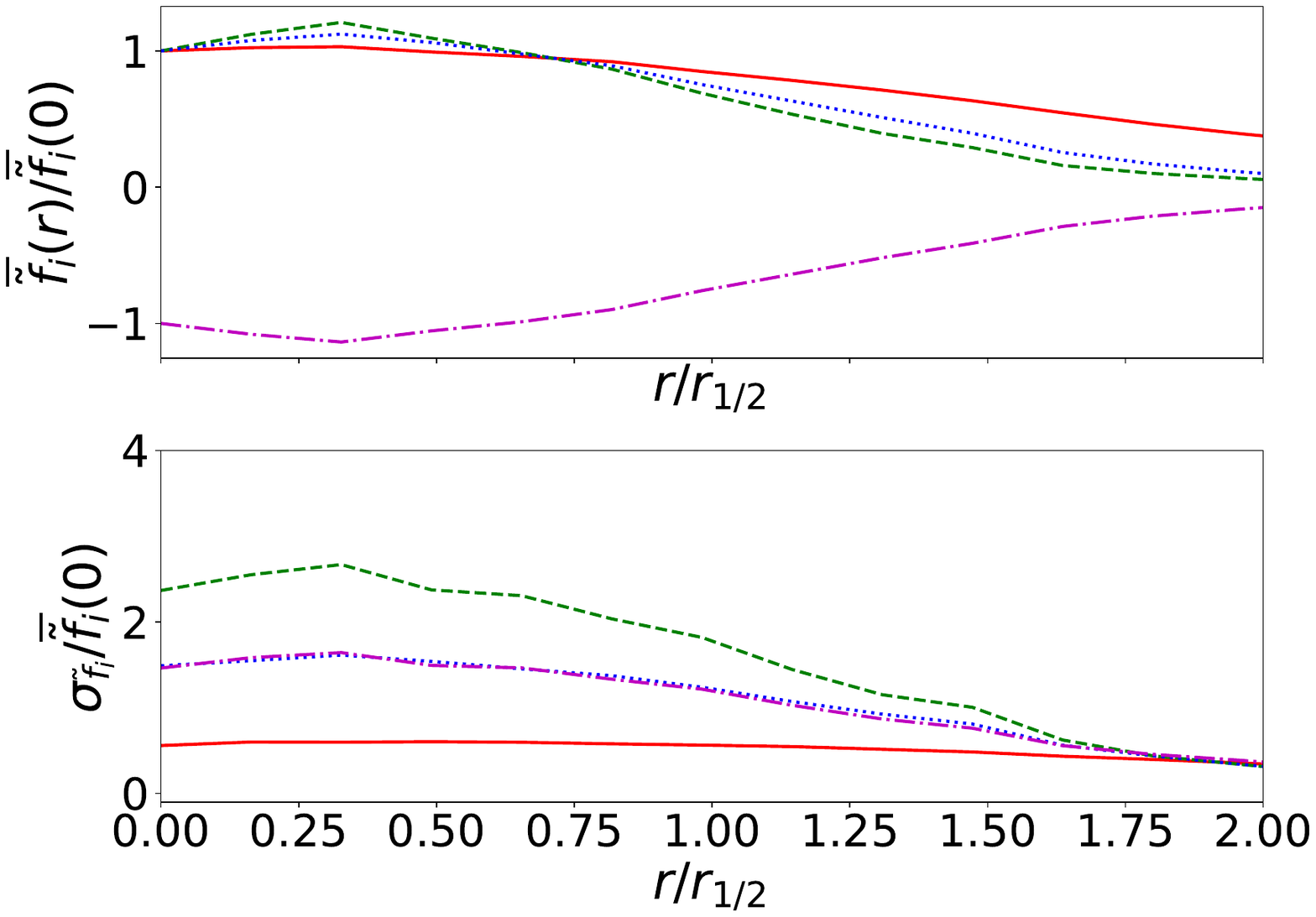}}
        \hfill
        \subfloat[SIM 2\label{fig:ts_we}]{\includegraphics[width=0.5\linewidth]{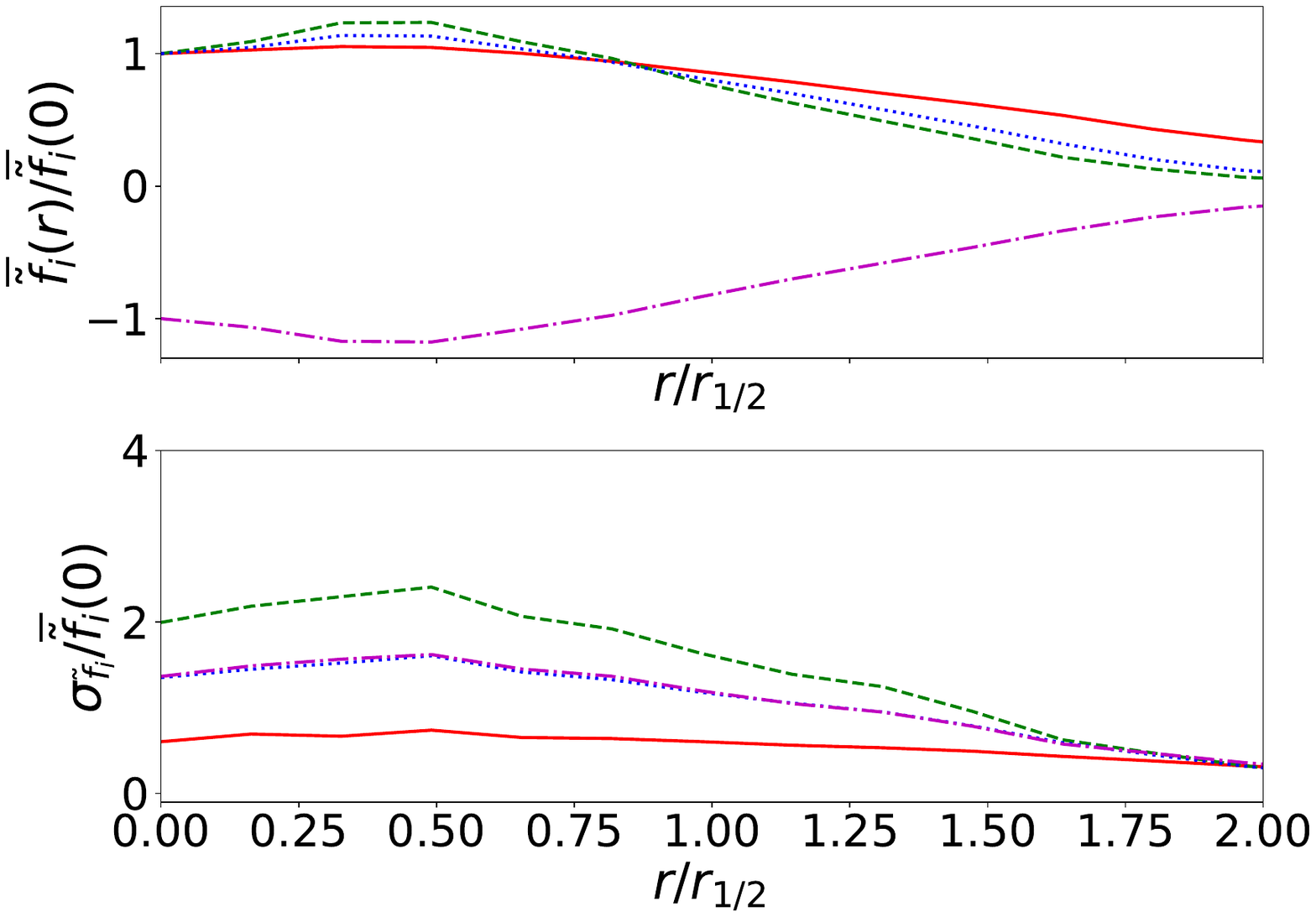}}

  \caption{The top panel depicts the radial distribution of the averaged inverse breakup timescale, $\tilde{t}_i = \widetilde{S}_{b,i}/\tilde{n_i}$ normalized by its centerline value while the bottom panel depicts the normalized standard deviation for (a) SIM 1 and (b) SIM 2.The lines are $d = 14\;\mu \mbox{m}$ (\protect\redline), $d = 100\;\mu \mbox{m}$ (\protect\greendline), $d = 550\;\mu \mbox{m}$ (\protect\blueddline) and $d = 3000\;\mu \mbox{m}$ (\protect\magdline) at $z/D_{J}= 70$.}
    \label{fig:ts_320}
\end{figure}

The breakup source term $\widetilde{S}_{b,i}$ normalized by the droplet concentration $\tilde{n}_i$ provides quantification of the inverse timescale for the breakup :
\begin{equation}
    \tilde{f}_i = \left\langle\frac{\widetilde{S}_{b,i}}{\tilde{n}_i}\right\rangle.
\end{equation}{}
This ratio can be interpreted as an inverse time-scale for droplet breakup to appreciably change the concentration of a particular size. Figure \ref{fig:ts_z} depicts the near nozzle evolution of five representative droplet sizes as a function of distance from the nozzle. We can see that the values are high near the nozzle exit where the breakup is rapid. The negative sign denotes that the $d=3\;mm$
and $d = 2261 \;\mu\mbox{m}$ droplets, on  average, act as sources for the smaller ones. A value of $\overline{\tilde{f}} =-60$ at the centerline means that it takes $1/60$ of a second for the local breakup to appreciably change the concentration of that droplet size. The values of the inverse timescale are also non monotonic with respect to droplet size. For instance, the   time-scale for $d=100\;\mu\mbox{m}$ droplet size is more rapid than the $d= 14\;\mu \mbox{m}$ droplet size. We can see that the inverse time scale increases for SIM 2 indicating that at larger Weber number the concentration changes more quickly, on a shorter time-scale.
The radial profiles for the mean inverse time scale and its variability at $z/D_{J} =70$ are shown in figure \ref{fig:ts_320}. From the top panel of figure \ref{fig:ts_src} we observe that the breakup is most rapid slightly off centre of the jet and then decays towards the edge of the jet. We can see that there is a high variability across the jet width, reflecting the underlying intermittency of the turbulent flow. 
Although an increase in Weber number results in a higher inverse breakup time-scale, the normalized radial profiles of the mean and variance appear relatively unchanged.  

This variability analysis from LES can be used as a tool to determine inherent fluctuations due to turbulence in measured quantities characterizing the droplet size distribution. Droplet Weber number, although having a significant effect on the average distribution of various quantities, leaves the normalized radial profiles of the mean and standard deviations relatively unchanged.

\section{Conclusions}\label{sec:conclusions}

Accurate prediction of the droplet size distribution in a turbulent flow is paramount in understanding the dynamics of numerous multiphase processes. We have applied a population balance model coupled with LES to study the evolution of oil droplets in an axisymmetric turbulent jet. In order to provide more realistic injection conditions for coarse-grid LES, we develop a 1D ODE model that predicts the evolution of the dispersed phase at the centerline turbulent jet by incorporating effects of droplet breakup and turbulent transport. The model is based on classical turbulent jet theory and is validated with experiments of \citet{Brandvik2013}, obtaining good agreement. The 1D ODE model is then used to provide an injection condition for a coarse grid LES of a turbulent jet. 
We perform two simulations with distinct Weber numbers to study surface tension effects on the evolution of the droplet size distributions.  The axial profiles of the individual droplet fields show interesting differences in the breakup dominated zone, exhibiting a size dependent decay rate. The radial profiles for the velocity and total concentration are, to a good approximation, self-similar and show good agreement with DNS results. We observe an off-axis peak for the total variance, similar to that observed in the evolution of a passive scalar. The droplet size distribution from the LES showed excellent agreement with both experimental data and the 1D ODE model. Additionally, LES is able to quantify new properties of the size distribution generated due to the inherent variability of turbulence. We quantify the radial profiles of the mean and variance of the characteristic diameter, total area available for surface reactions, and the normalized breakup source terms. In accordance with numerous experiments, we observe that the Sauter mean diameter, defined as the volume to surface area ratio of the distribution, decreases with increasing Weber number. This can be attributed to increased breakup of larger droplets resulting in a steeper slope in the small-scale size range of the droplet size distribution. Although demonstrating a significant effect on the averaged droplet size distribution, the Weber number has minimal effect on the radial profiles of the normalized standard deviations of key quantities.   

\section*{Acknowledgements}

The authors thank D. Yang, M. Chamecki, T. Chor, J. Katz, X. Xue, R. Ni,  for useful conversations and insights. This research was made possible by a grant from the Gulf of Mexico Research Initiative. Computational resources were provided by the Maryland Advanced Research Computing Center (MARCC). Data are publicly available through Gulf of Mexico Research Inititative Information and Data Cooperative (GRIIDC) at https://data.gulfresearchinitiative.org (doi:10.7266/YNHKDHS4).

\bibliography{Oil_jet.bib}% Produces the bibliography via BibTeX.

\end{document}